\documentclass{ar-1col}

\pdfoutput=1

\usepackage[numbers,sort,compress]{natbib}
\usepackage{url}
\usepackage{amsmath}
\usepackage{amssymb}
\usepackage{xcolor}
\usepackage[colorlinks,linkcolor=blue,citecolor=blue]{hyperref}
\usepackage[english]{babel}

\hypersetup{colorlinks=true, linkcolor=blue, urlcolor=blue}

\jname{Ann. Rev. Nuc. Part. Sci.}
\jvol{71}
\jyear{2021}
\doi{10.1146/annurev-nucl-102920-050505}

\begin{document}

\markboth{Tamborra \& Shalgar}{Flavor Evolution of Dense Neutrino Gas}

\title{New Developments in Flavor Evolution of a Dense Neutrino Gas}

\author{Irene Tamborra$^1$ and Shashank Shalgar$^1$
\affil{$^1$Niels Bohr International Academy \& DARK, Niels Bohr Institute,\\University of Copenhagen, 2100 Copenhagen, Denmark, email: tamborra@nbi.ku.dk}}

\begin{abstract}
Neutrino-neutrino refraction dominates the flavor evolution in core-collapse supernovae, neutron-star mergers, and  the early universe. Ordinary neutrino flavor conversion  develops on timescales determined by the vacuum oscillation frequency. However, when the neutrino density is large enough, collective flavor conversion  may arise because of pairwise neutrino scattering. Pairwise conversion is deemed to be fast as  it is expected to occur on timescales  that depend on the neutrino-neutrino interaction energy (i.e., on the neutrino number density) and is regulated by the  angular distributions of electron neutrinos and antineutrinos. The  enigmatic phenomenon of fast pairwise conversion has been overlooked for a long time. However, because of the fast conversion rate, pairwise conversion  may possibly occur in the proximity of the neutrino decoupling region with yet to be understood implications for the hydrodynamics of astrophysical sources and the synthesis of the heavy elements. We review the physics of this fascinating phenomenon  and its implications for neutrino-dense sources. 

\end{abstract}

\maketitle

{\hypersetup{hidelinks}

\tableofcontents
}


\section{INTRODUCTION}
Neutrinos are among the  most abundant particles  in our universe, playing  a pivotal role in a variety of astrophysical environments, ranging from the sun to the most energetic transients~\cite{Vitagliano:2019yzm}.
Neutrinos interact weakly and have the unique property of changing their flavor while propagating. In neutrino dense environments, such as core-collapse supernovae, neutron-star mergers and the early universe,  the flavor evolution is  vastly affected by  the interactions of neutrinos  among themselves~\cite{Duan:2010bg,Mirizzi:2015eza,Chakraborty:2016yeg}, in addition to  ordinary neutrino interactions with matter. Neutrino-matter interactions,  under certain circumstances, can lead to the resonant conversion of (anti)neutrinos, the so called Mikheyev-Smirnov-Wolfenstein (MSW) resonance~\cite{Mikheev:1986if,1985YaFiz..42.1441M,1978PhRvD..17.2369W}. 

\begin{marginnote}[]
\entry{Mikheyev-Smirnov-Wolfenstein (MSW) resonance}{Enhanced flavor conversion of (anti)neutrinos in matter.}
\end{marginnote}

It is especially relevant to grasp the  physics of  neutrino-neutrino interaction as it may give insights into astrophysics, fundamental neutrino properties, and non-standard scenarios involving neutrinos.
The forward coherent scattering of neutrinos among themselves~\cite{1987ApJ...322..795F,Notzold:1987ik},  misleadingly called ``neutrino self-interaction,''  crucially differs from  neutrino interaction with matter.  
 Neutrino-neutrino interaction leads to a non-linear evolution with a positive feedback  of the neutrino field onto itself~\cite{Pantaleone:1992eq,Pantaleone:1992xh}. As a consequence, the  flavor evolution develops a collective nature  that manifests itself  through the coupling  of all  momentum modes.

\begin{figure}[t]
\includegraphics[width=4.in]{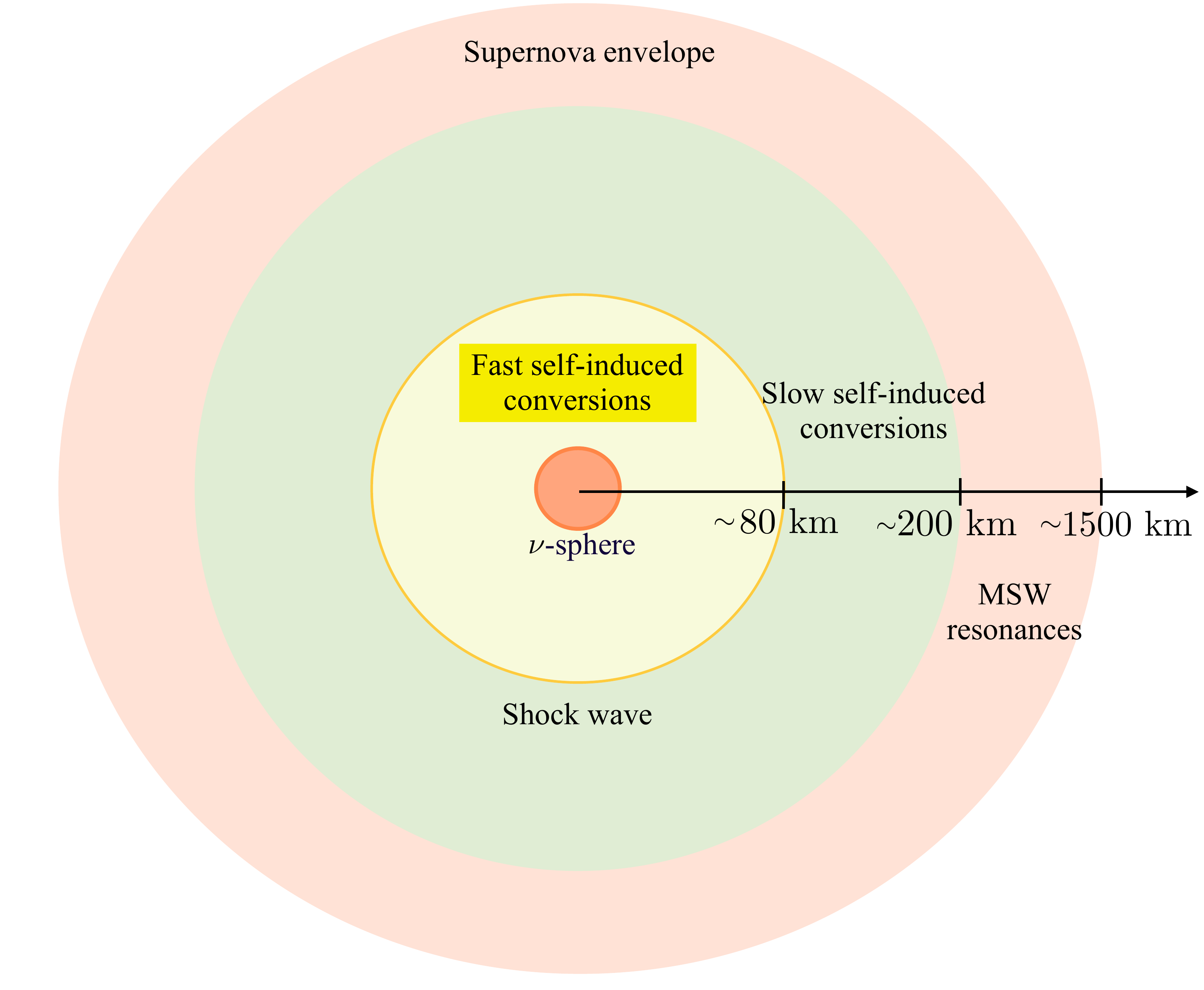}
\caption{Schematic representation of the neutrino flavor conversion  regions in the  envelope of a core-collapse supernova; indicative distances from the supernova core are displayed. According to the classical picture, MSW conversions and $\nu$--$\nu$ interactions in the slow regime are expected to occur beyond the shock radius (in the red- and green-shadowed areas, respectively). However, if fast pairwise conversions should take place, they might occur in the proximity of the supernova core  ($\nu$-sphere) and within the shock wave radius (yellow-shadowed area), possibly affecting the supernova fate itself.}
\label{fig:SNsketch}
\end{figure}

The physics of neutrino-neutrino interaction has mostly been explored in the context of core-collapse supernovae~\cite{Mirizzi:2015eza,Janka:2006fh,Janka:2017vcp,Burrows:2012ew,Burrows:2020qrp}, originating from the death of massive stars. It was pointed out that neutrino-neutrino interaction leads to ``slow'' conversions~\cite{Mirizzi:2015eza,Duan:2010bg}  occurring on a timescale that is determined by the vacuum oscillation frequency, $\omega = \Delta m^2/2E \simeq 6.3\ \mathrm{km}^{-1}/E[\mathrm{MeV}]$, where $\Delta m^2$ is the largest squared mass difference of neutrinos and $E \simeq \mathcal{O}(15)$~MeV is the typical neutrino energy. The neutrino-neutrino interaction strength, $\mu = \sqrt{2} G_{\rm F} (n_\nu + {n}_{\bar\nu})$, is determined by the Fermi constant $G_{\rm F}$ and the (anti)neutrino number density for all six flavors $n_\nu ({n}_{\bar\nu})$; $\mu$ scales with the distance from the neutrino emission surface, but it is $\mathcal{O}(10^5)$~km$^{-1}$ in the proximity of the neutrino decoupling region.  In the widely investigated spherically symmetric supernova model,  ``slow''  neutrino-neutrino interactions are expected to be relevant away from the spatial region where (anti)neutrinos decouple, as schematically shown in \textbf{Fig.~\ref{fig:SNsketch}}, and they may lead to characteristic signatures in the (anti)neutrino energy distributions such as the so-called spectral splits, i.e.~certain energy modes swap their flavor content according to the neutrino mass ordering~\cite{Duan:2006an,Fogli:2007bk}. For a  long time,  spectral splits  were considered  to be a  characteristic imprint of neutrino-neutrino interactions.  

This neat picture became more complex in the past decade. In fact, given the challenges related to the numerical implementation of the neutrino feedback onto itself, many symmetries were initially imposed  to solve the equations of motion. It was soon realized that these  symmetries could be  broken spontaneously,  affect the flavor stability conditions, and trigger flavor conversions at higher densities~\cite{Raffelt:2013rqa,Duan:2014gfa,Abbar:2015mca,Mirizzi:2015fva,Cirigliano:2017hmk}. In addition, it became clear that flavor conversions are not only affected by the energy distribution, but also by the angular distribution of (anti)neutrinos~\cite{Mirizzi:2015eza,Chakraborty:2016yeg,Fogli:2007bk,Duan:2006an}.

The  rough understanding of the phenomenology of neutrino-neutrino interactions has been shaken by the realization that, when the neutrino density is large enough, neutral current interactions of the type $\nu_e(\vec{p})+\nu_x(\vec{q}) \leftrightarrow \nu_x(\vec{p})+\nu_e(\vec{q})$ and $\nu_e(\vec{p})+\bar\nu_e(\vec{q}) \leftrightarrow \nu_x(\vec{p})+\bar\nu_x(\vec{q})$, with $\nu_x = 
\nu_\mu$ or $\nu_\tau$ and $\vec{p} (\vec{q})$ the neutrino momentum,  are not negligible~\cite{Sawyer:2005jk,Sawyer:2008zs}. Pairwise interactions can, strictly speaking, occur even if the 
vacuum term is zero (differently from slow conversions), they preserve the net neutrino flavor, but they can  modify the subsequent charged-current interactions.  
The characteristic scale of  pairwise conversions is  the neutrino-neutrino interaction strength,  $\mu$. Flavor conversions triggered by  pairwise neutrino scattering are deemed ``fast,'' in the sense that they develop on  timescales orders of magnitude smaller than the ones of slow neutrino-neutrino interactions.  Only recently, it has been appreciated that, if fast pairwise conversions are at play, they could drastically affect the flavor content in the proximity of the neutrino decoupling region (see \textbf{Fig.~\ref{fig:SNsketch}}) with possible  significant implications on the source dynamics and nucleosynthesis~\cite{Sawyer:2015dsa,Sawyer:2005jk,Sawyer:2008zs,Chakraborty:2016lct,Chakraborty:2016yeg,Izaguirre:2016gsx}.

\begin{marginnote}[]
\entry{Pairwise neutrino scattering}{Neutral current interactions between pairs of neutrinos with an exchange of momentum, but not lepton number.}
\entry{Slow flavor conversions}{Develop on  timescales mainly determined by the neutrino vacuum frequency.}
\entry{Fast flavor conversions}{Develop on  timescales dictated by the neutrino-neutrino interaction strength.}
\end{marginnote}

Despite intense theoretical work, our understanding of  the role of neutrino-neutrino interaction in  dense media,  and especially of fast conversion, is still very approximate.  A numerical solution  entails solving a seven-dimensional transport problem (involving time, three spatial coordinates, and three momentum coordinates) with characteristic quantities spanning  several orders of magnitude.  As such, a realistic and self-consistent study is out of reach given the computational methods and resources currently available.  On the other hand, one of the main goals is to predict whether  maximal or negligible flavor mixing  could be achieved, in order to grasp the implications of neutrino-neutrino interaction in astrophysics and cosmology.

In this review, we  focus on simple examples to outline the phenomenology of fast pairwise conversion, which is 
  the most recent and less explored phenomenon characteristic of  neutrino propagation in dense media. 
  The paper is organized as follows. In Sec.~\ref{sec:oscillations}, the neutrino equations of motion are introduced together with the linear stability analysis, and the conditions possibly leading to the growth of fast flavor instabilities. Section~\ref{sec:pheno} focuses on the phenomenology of fast pairwise conversions; their dependence on the neutrino vacuum frequency, electron lepton number, neutrino advection, and collisions is outlined. The possible occurrence of fast pairwise conversions in core-collapse supernovae, compact binary mergers, and the early universe is discussed in Secs.~\ref{sec:SN}, \ref{sec:NSM}, and \ref{sec:EU}, respectively. Finally, summary and outlook follow.

\section{NEUTRINO FLAVOR CONVERSIONS}\label{sec:oscillations}
In this section, we introduce the  equations of motion for neutrinos in the mean-field approximation, although many-body descriptions can also be adopted~\cite{Friedland:2003eh,Friedland:2003dv,Balantekin:2006tg,Pehlivan:2010zz,Pehlivan:2011hp,Birol:2018qhx,Patwardhan:2019zta,Cervia:2019res,Rrapaj:2019pxz}. The onset of flavor conversions can be examined through a linear stability analysis applied to the linearized equations of motion.  In order to grasp the development of fast pairwise conversions, we introduce the linear stability analysis technique, discuss under which conditions flavor conversions develop, and classify the kind of flavor instability.

\subsection{Equations of motion}
The neutrino evolution in dense media is described in terms of the neutrino  flavor-field represented  by the Wigner-transformed density matrix in the flavor space $\rho(t,\vec{x},\vec{p})$---and $\bar\rho(t,\vec{x},\vec{p})$ for antineutrinos---expressed as a function of time ($t$), location ($\vec{x}$), and momentum ($\vec{p}$). 
The  density matrix in the flavor basis, $\rho(t,\vec{x},\vec{p})$, has elements that are expectation values of bilinear creation and annihilation operators $\langle a_i^\dag a_j \rangle$, where $i$ and $j$ are flavor indices. 
The diagonal elements of the density matrix are the occupation numbers, while the off-diagonal ones describe the flavor correlations. The seven-dimensional phase space is not tractable numerically, and it is usually reduced through symmetry assumptions. For example, one often looks for stationary solutions only depending  on the radial coordinate,  energy, and zenith angle; hence, reducing the problem to three dimensions.

For simplicity, in what follows, we rely on a two flavor approximation in the weak-interaction basis: $(\nu_e,\nu_x)$,  where $x$ stands for $\mu$, $\tau$, or a linear combination of the two.
The equations of motion that describe the  flavor evolution of ultra-relativistic neutrinos are~\cite{Sigl:1992fn,Cardall:2007zw}
\begin{equation}\label{eq:EOM1}
 \left(\frac{\partial}{\partial t} + \vec{v} \cdot \vec{\nabla}_{x} + \vec{F}\cdot \vec{\nabla}_{p}\right) \rho(\vec{x},\vec{p}, t)= -\imath [H(\vec{x},\vec{p}, t),\rho(\vec{x},\vec{p}, t)] +  \mathcal{C}\left(\rho(\vec{x},\vec{p}, t),\bar{\rho}(\vec{x},\vec{p}, t)\right)\ ,
\end{equation}
\begin{equation}\label{eq:EOM2}
 \left(\frac{\partial}{\partial t} + \vec{v} \cdot \vec{\nabla}_{x} + \vec{F}\cdot \vec{\nabla}_{p}\right) \bar{\rho}(\vec{x},\vec{p}, t) = - \imath [\bar{H}(\vec{x},\vec{p}, t),\bar\rho(\vec{x},\vec{p}, t)]+  \bar{\mathcal{C}}\left(\rho(\vec{x},\vec{p}, t),\bar{\rho}(\vec{x},\vec{p}, t)\right)\ ,
\end{equation}
where the advective term, $\vec{v} \cdot \vec{\nabla}_{x}$, depends on the velocity of the (anti)neutrino field $\vec{v}$ and it  is non-zero in the presence of  spatial variations.  The term proportional to the force $\vec{F}$ is mathematically similar to the advective term; however, it depends on the gradient in the momentum space~\cite{Stirner:2018ojk}. This term takes into account the change in energy, direction, or both during propagation;  e.g.,~it could  result from an external force, such as the gravitational force that can bend the neutrino trajectory. 
On the right hand side, $[\dots,\dots]$  denotes the commutator of two matrices.  The term $\mathcal{C} (\bar{\mathcal{C}})$ schematically represents the collision term, which takes into account the non-forward scattering of neutrinos with the medium or other neutrinos. 

The Hamiltonian that describes the temporal evolution of the density matrix  is made of three terms: the vacuum term, the matter term, and the neutrino-neutrino term,
\begin{eqnarray}
H(\vec{x},\vec{p}, t)=H_{\mathrm{vac}}+H_{\mathrm{mat}}+H_{\nu\nu}\ \ \mathrm{and}\ \ 
\bar{H}(\vec{x},\vec{p}, t)=-H_{\mathrm{vac}}+H_{\mathrm{mat}}+H_{\nu\nu}\ .
\end{eqnarray}
The various terms of the Hamiltonian have the following form:
\begin{subequations} \label{eq:hamiltonian}
\begin{eqnarray}
        H_{\textrm{vac}} &=& 
\frac{\omega}{2} \left(\begin{matrix}
       -\cos 2 \theta_{\textrm{V}} & \sin 2 \theta_{\textrm{V}} \\
        \sin 2 \theta_{\textrm{V}} & \cos 2 \theta_{\textrm{V}}
\end{matrix}\right)\ , \\
        H_{\textrm{mat}} &=& 
        \left(\begin{matrix}
                \sqrt{2} G_{\textrm{F}} n_{e} & 0 \\
                0 & 0
        \end{matrix}\right)\ , \\
        H_{\nu\nu} &=& \mu \int d\vec{p^{\prime}}[\rho(\vec{p^\prime})-\bar{\rho}(\vec{p^\prime})]
        \left(1- \vec{v}\cdot\vec{v^{\prime}}\right)\ .
        \end{eqnarray}
\end{subequations}
The vacuum frequency is $\omega =  \Delta m^2/2E$  (with $E$ being the neutrino energy and $\Delta m^2 >0$ for normal mass ordering and $\Delta m^2 <0$  for inverted ordering), $\theta_{\textrm{V}}$ is the vacuum mixing angle, $n_{e}$ is the effective number density of electrons in the medium, and $\mu = \sqrt{2} G_F (n_{\nu}+{n}_{\bar\nu})$ is the strength of the $\nu$--$\nu$ potential which is proportional to the (anti)neutrino number density for all flavors, $n_{\nu}$ (${n}_{\bar\nu}$). Note that we have chosen a basis such that $H_{\textrm{vac}}$  and $H_{\textrm{mat}}$  have the same (opposite)  sign for (anti)neutrinos~\cite{Duan:2006an}. Moreover, we have dropped the terms proportional to the identity matrix in the Hamiltonian because they only contribute to overall phases;  these include the neutral current interactions in the matter term, and a term proportional to the trace in the $\nu$--$\nu$ interaction Hamiltonian. However, these terms may become non-diagonal, for example, when the Hamiltonian is expanded to include physics beyond the Standard Model, see e.g.~Refs.~\cite{Nunokawa:1997ct,Stapleford:2016jgz}, but such scenarios are not considered here. 
 We have also excluded non-local terms from the Hamiltonian where the lepton asymmetry is large, which are not of significance in compact astrophysical objects,  but become important in the early universe~\cite{Wong:2002fa,Dolgov:2002ab,Johns:2016enc}.

One of the most significant developments in the field of neutrino flavor conversion physics was the observation by Mikheyev, Smirnov, and Wolfenstein  that, under certain conditions,  the diagonal components of $H_{\textrm{vac}}$ and $H_{\textrm{mat}}$ can cancel each other, leading to the MSW resonant conversion of (anti)neutrinos~\cite{Mikheev:1986if,1985YaFiz..42.1441M,1978PhRvD..17.2369W}. 
Similarly, the  non-linear $\nu$--$\nu$ term in the Hamiltonian  can be responsible for non-negligible flavor conversion, since neutrino self-interactions can trigger  an exponential growth of the off-diagonal terms of the Hamiltonian in time. However, a characteristic feature of $\nu$--$\nu$ interaction is that it does not only depend on the  energy distributions  (like MSW oscillations), but also on the angular distributions of neutrinos and antineutrinos.

\subsection{Linear stability analysis}\label{sec:LSA}
The phenomenology of neutrino-neutrino interactions is very counter-intuitive to grasp because of the non-linear nature of the problem.  Reference~\cite{Banerjee:2011fj} proposed to extend  the linear stability analysis technique, widely adopted  to understand whether a system has unstable solutions in many areas of Physics, to the  linearized equations of motion of neutrinos. The linear stability analysis allows to predict under which conditions 
$\nu$--$\nu$ interactions  lead to the growth of flavor instabilities (i.e., possibly to flavor mixing).

For the sake of simplicity, let us consider a homogeneous gas with no collisions and no external force. In addition, let us  assume that  neutrinos are mono-energetic and the momentum depends on two angles, the polar angle $\theta$ and the azimuthal angle $\phi$. 
For each given tuple $(\theta,\phi)$, the density matrix encoding the flavor information of (anti)neutrinos traveling in that direction at a given time can be defined as
\begin{eqnarray}
\rho(\vec{x},\theta,\phi,t) &=&
\begin{pmatrix}
\rho_{ee}(\vec{x},\theta,\phi,t=0) - \Delta(\vec{x},\theta,\phi,t) & \epsilon(\vec{x},\theta,\phi,t) \cr
\epsilon^{*}(\vec{x},\theta,\phi,t) & \rho_{xx}(\vec{x},\theta,\phi,t=0) + \Delta(\vec{x},\theta,\phi,t)
\end{pmatrix}
\\
\bar{\rho}(\vec{x},\theta,\phi,t) &=&
\begin{pmatrix}
\bar{\rho}_{ee}(\vec{x},\theta,\phi,t=0) - \bar{\Delta}(\vec{x},\theta,\phi,t) & \bar{\epsilon}(\vec{x},\theta,\phi,t) \cr
\bar{\epsilon}^{*}(\vec{x},\theta,\phi,t) & \bar{\rho}_{xx}(\vec{x},\theta,\phi,t=0) + \bar{\Delta}(\vec{x},\theta,\phi,t)
\end{pmatrix}
\end{eqnarray}
where, {$|\epsilon(\vec{x},\theta,\phi,t)|\ll |\rho_{ee}(\vec{x},\theta,\phi,t)-\rho_{xx}(\vec{x},\theta,\phi,t)|$ and $|\bar{\epsilon}(\vec{x},\theta,\phi,t)|\ll |\bar{\rho}_{ee}(\vec{x},\theta,\phi,t)-\bar{\rho}_{xx}(\vec{x},\theta,\phi,t)|$.}

The equations of motion (Eqs.~\ref{eq:EOM1} and \ref{eq:EOM2}) can be expanded in series in  $\epsilon$ (and $\bar{\epsilon}$), and it can  easily be seen that the evolution of $\Delta(\vec{x},\theta,\phi,t)$ goes like $\mathcal{O}(\epsilon^{2})$. By focusing on the leading order (and therefore neglecting $\Delta$), the equation of motion for $\epsilon(\vec{x},\theta,\phi,t)$ is:
\begin{eqnarray}
 \left(\frac{\partial}{\partial t}+\vec{v}\cdot\vec{\nabla}\right) \epsilon(\vec{x},\theta,\phi,t)= - \imath \int d\cos\theta^{\prime} d\phi^{\prime} U(\vec{x}, \theta,\phi,\theta^{\prime},\phi^{\prime},t) \epsilon(\vec{x}, \theta^{\prime},\phi^{\prime},t)\ ;
\label{lin:eom}
\end{eqnarray}
a similar equation holds for $\bar{\epsilon}$. The  evolution of the equation above suggests that   the off-diagonal terms of the density matrix may rapidly become  comparable to the diagonal terms; when this happens, the evolution of the off-diagonal terms is no longer exponential; this is called the {\it non-linear regime} in the literature and  we have to rely on numerical techniques to gain any insight. 

If the matrix $U$ has a complex eigenvalue, an exponential growth of the corresponding eigenvector will occur, which in turn contributes to the off-diagonal component of the Hamiltonian triggering the onset of neutrino flavor conversions. The exponential growth of a particular eigenvector is called {\it flavor instability}~\cite{Banerjee:2011fj,Sarikas:2011am}. Although the existence of a flavor instability is a promising indicator of flavor conversions, significant flavor conversion is not guaranteed, if a flavor instability exists. 

\begin{marginnote}[]
\entry{Flavor instability}{Exponential growth of the modulus of the off-diagonal terms of the density matrix in the linearized equations of motion. Its existence is a necessary, but not sufficient, condition for flavor conversions to occur. }
\end{marginnote}

If we assume that $\epsilon$ and $U$ do not depend on $\phi$,  the linearized equations of motion  for neutrinos and antineutrinos become 
\begin{eqnarray}
\label{eq:lin1a}
\imath\left(\frac{\partial}{\partial t}+\vec{v}\cdot\vec{\nabla}\right)\epsilon(\vec{x},\theta,t) &=& (H_{ee}-H_{xx}) \epsilon(\vec{x},\theta,t) \nonumber \\
&+& (\rho_{xx}-\rho_{ee}) \mu \int d\cos\theta^{\prime} [\epsilon(\vec{x},\theta^{\prime}, t) - \bar{\epsilon}(\vec{x}, \theta^{\prime}, t)] (1-\vec{v} \cdot \vec{v}^{\prime})\ , \\
\imath\left(\frac{\partial}{\partial t} + \vec{v}\cdot\vec{\nabla} \right)\bar\epsilon(\vec{x}, \theta,t) &=& (\bar{H}_{ee}-\bar{H}_{xx}) \bar{\epsilon}(\vec{x}, \theta,t) \nonumber \\
&+& (\rho_{xx}-\rho_{ee}) \mu \int d\cos\theta^{\prime}  [\epsilon(\vec{x}, \theta^{\prime}, t) - \bar{\epsilon}(\vec{x}, \theta^{\prime}, t)] (1-\vec{v} \cdot \vec{v}^{\prime})\ ,
\label{eq:lin1b}
\end{eqnarray}
where the terms $\rho_{ij}$ and $H_{ij}$ denote the elements of the $2\times2$ density matrix and the Hamiltonian, respectively. 
 The evolution of neutrinos and antineutrinos is coupled when  the $\nu$--$\nu$ term dominates; hence, the flavor instabilities of (anti)neutrinos grow at the same rate~\cite{Banerjee:2011fj}:
\begin{eqnarray}
\label{eq:lin2a}
\epsilon(\vec{x},\theta,t) = Q_{\theta} e^{-\imath (\Omega t-\vec{k}\cdot\vec{x})}\  \mathrm{and}\ 
\bar{\epsilon}(\vec{x},\theta,t) = \bar{Q}_{\theta} e^{-\imath (\Omega t-\vec{k}\cdot\vec{x})}\ .
\end{eqnarray}
When $\Omega$ and $\vec{k}$ have non-zero imaginary solutions, the flavor instability grows exponentially.
In the homogeneous case ($\vec{k}=0$),   the so-called temporal instabilities occur for $\mathrm{Im}(\Omega)\neq 0$ and grow in time exponentially; while if $\Omega=0$, the so-called spatial instabilities occur for $\mathrm{Im}(\vec{k})\neq 0$; in this case, the flavor instabilities grow in space~\cite{Chakraborty:2016yeg}. 

In the homogeneous case ($\vec{k}=0$), substituting Eq.~\ref{eq:lin2a}  in Eqs.~\ref{eq:lin1a} and \ref{eq:lin1b}, we obtain:
\begin{eqnarray}
\left[\Omega - (H_{ee}-H_{xx})\right] Q_{\theta} e^{-\imath \Omega t} = (\rho_{xx}-\rho_{ee}) \mu \int d\cos\theta^{\prime} (Q_{\theta^{\prime}} - \bar{Q}_{\theta^{\prime}}) e^{-\imath \Omega t} (1-\vec{v} \cdot \vec{v}^{\prime})\ , \\
\left[\Omega - (\bar{H}_{ee}-\bar{H}_{xx})\right] \bar{Q}_{\theta} e^{-\imath \Omega t} = (\bar{\rho}_{xx}-\bar{\rho}_{ee}) \mu \int d\cos\theta^{\prime} (Q_{\theta^{\prime}} - \bar{Q}_{\theta^{\prime}}) e^{-\imath\Omega t} (1-\vec{v} \cdot \vec{v}^{\prime})\ .
\label{lin3}
\end{eqnarray}
The latter equations  allow to express $Q_{\theta}$ and $\bar{Q}_{\theta}$ in the following parametric form:
\begin{eqnarray}
Q_{\theta} = (\rho_{xx}-\rho_{ee}) \frac{(a-b \cos\theta)}{\Omega - (H_{ee}-H_{xx})} \ \mathrm{and}\ 
\bar{Q}_{\theta} = (\bar{\rho}_{xx}-\bar{\rho}_{ee}) \frac{(a-b \cos\theta)}{\Omega - (\bar{H}_{ee}-\bar{H}_{xx})}\ , 
\label{lin4}
\end{eqnarray}
which depends on the   parameters, $a$ and $b$, and $\Omega$. From here, the eigenvalue $\Omega$ can be computed by solving the following equation~\cite{Chakraborty:2016yeg}:
\begin{eqnarray}
\begin{vmatrix}
I[1]-1 & -I[\cos\theta] \cr
I[\cos\theta] & -I[\cos^{2}\theta]-1
\end{vmatrix} = 0\ ,
\label{lin5}
\end{eqnarray} 
where
\begin{eqnarray}
I[f(\theta)] = \int d\cos\theta f(\theta) \left[\frac{(\rho_{ee}-\rho_{xx})}{\Omega-(H_{ee}-H_{xx})}-\frac{(\bar{\rho}_{ee}-\bar{\rho}_{xx})}{\Omega-(\bar{H}_{ee}-\bar{H}_{xx})}   \right]\ .
\label{lin6}
\end{eqnarray}
Equation~\ref{lin5} can be solved analytically with respect to $\Omega$ for special classes of angular distributions of (anti)neutrinos, see e.g.~Refs.~\cite{Cirigliano:2017hmk,Chakraborty:2016lct}. 
Similarly, one can follow the same procedure and look for $\mathrm{Im}(\vec{k}) \neq 0$ when $\Omega=0$ or, more generally,  for imaginary solutions of $(\Omega,\vec{k})$. 

\begin{figure}
\includegraphics[width=\textwidth]{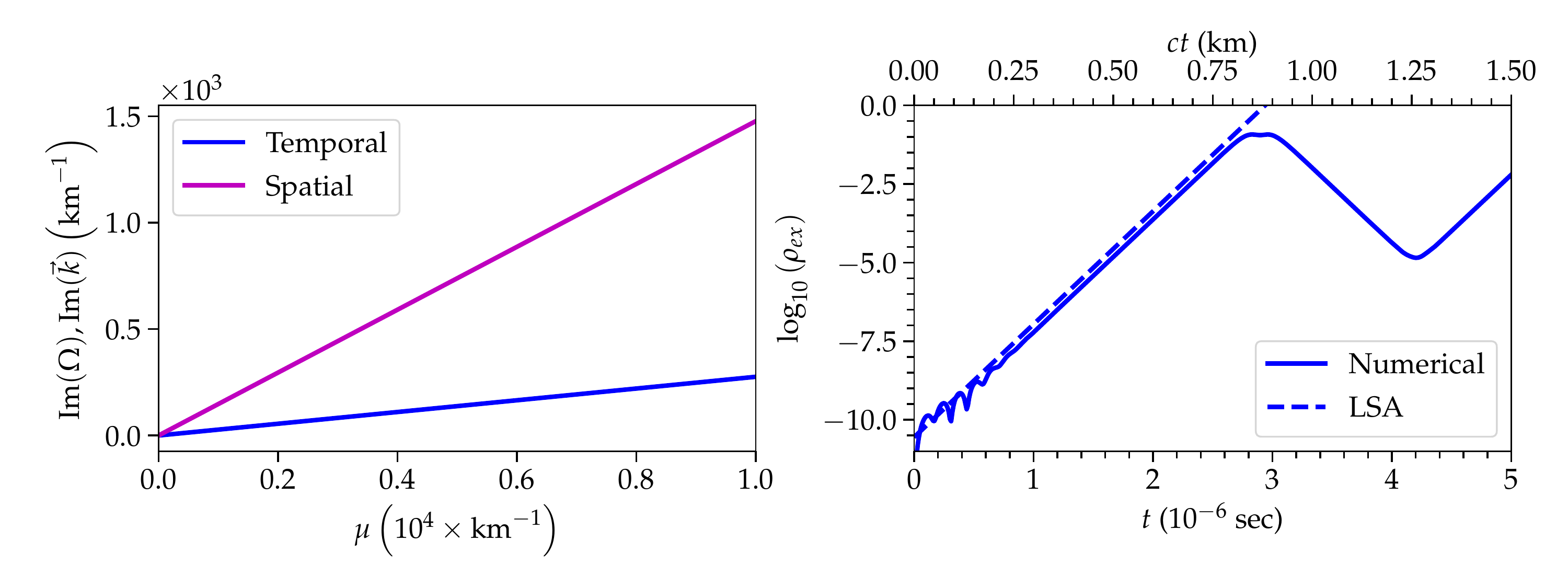}
\caption{Flavor instabilities predicted by the linear stability analysis. {\it Left:} Temporal ($\mathrm{Im}(\Omega) \neq 0$, blue line) and spatial ($\mathrm{Im}(\vec{k}) \neq 0$, magenta line) growth rate as a function of the neutrino self-interaction strength $\mu$ for Case 1 (see main text for details). The growth rate linearly depends  on $\mu$. 
{\it Right:} Temporal evolution of the angle-integrated off-diagonal term of the density matrix in the linear regime for $\vec{k}=0$ and $\mu=10^3$~km$^{-1}$ for Case 1. 
The prediction of the linear stability analysis (LSA, dashed line) is in excellent agreement with the numerical solution (solid line). 
}
\label{Fig1}
\end{figure}

In order to investigate the growth of flavor instabilities, we consider  a perfectly homogeneous neutrino gas, ignore the advective term $\vec{v}\cdot\vec{\nabla}_x$, and consider the following angular distributions (Case 1):
\begin{eqnarray}
\label{eq:case1}
\rho_{ee}(t=0) = 0.5 \ , \ \bar{\rho}_{ee}(t=0) = 
\begin{cases}
1 & \theta \in [0,\pi/3]  \\
0.25 & \theta \in [\pi/3,\pi]\ \end{cases},   \mathrm{and}\  \rho_{xx}(t=0)=\bar\rho_{xx}(t=0)=0 \ ,
   \label{rhodist}
\end{eqnarray}
such that $\int d\cos\theta \rho_{ee}(t=0)$ = 1. The blue line in the left panel {\bf Fig.~\ref{Fig1}} shows $\mathrm{Im}(\Omega)$  as a function of $\mu$, one can see that the flavor instability exists irrespective of the value of the self-interaction potential,  and grows linearly with $\mu$. A similar trend is found for  the magenta curve that represents  $\mathrm{Im}(\vec{k})$. Assuming $\vec{k}=0$, $\mu=10^3$~km$^{-1}$~\footnote{In order to speed up the numerical computations, the value of $\mu$ adopted in our  examples is smaller than the typical neutrino self-interaction strength in the decoupling region [$\mathcal{O}(10^5)$~km$^{-1}$]. A larger $\mu$ would lead to the development of flavor conversions on scales smaller than what shown here, but it would not affect the overall flavor phenomenology.}, $\Delta m^{2}=2.5\times 10^{-3}$~eV$^{2}$, $\theta_V = 10^{-6}$, and $E=10^{4}$~MeV, the right panel of {\bf Fig.~\ref{Fig1}} shows the temporal evolution of the angle-integrated off-diagonal term of the density matrix as from the predictions of the linear stability analysis and the numerical solution of the equations of motion (Eqs.~\ref{eq:EOM1} and \ref{eq:EOM2}). One can see that the linear stability analysis perfectly agrees with the numerical solution in the linear regime, but it cannot give any insight on the non-linear evolution of the density matrix. The linear stability analysis aims to predict whether the (anti)neutrino ensemble may develop flavor instabilities for given initial conditions. However, the linear stability analysis cannot predict the final flavor outcome as the latter is strongly affected by the non-linear regime of neutrino-neutrino interactions.

\subsection{Fast pairwise conversions}
While slow $\nu$--$\nu$ conversions become relevant relatively far away from the source (see {\bf Fig.~\ref{fig:SNsketch}}),  fast pairwise conversions are expected to occur at high densities, in the proximity of the decoupling region~\cite{Sawyer:2005jk,Sawyer:2008zs,Sawyer:2015dsa}. Being determined by pairwise scattering of (anti)neutrinos, they could  occur even for $\Delta m^2 =0$; hence, in the absence of external perturbations and for $\omega \rightarrow 0$, fast pairwise conversions are such  that the  net number
of electron, mu, or tau flavor carried by neutrinos is separately conserved, in addition to the total lepton number.

The linear stability analysis predicts that fast pairwise conversions of neutrinos should arise, for an arbitrarily large self-interaction potential,  in a neutrino gas where the number density of $\bar\nu_e$ is the same as the one of $\nu_e$ along some direction, but not other directions; i.e.,~there is at least one angle where a crossing between the angular distributions of $\nu_e$'s and $\bar\nu_e$'s  occurs (ELN crossing, temporal instability) or there is a non-negligible backward flux of neutrinos (spatial instability)~\cite{Chakraborty:2016yeg,Izaguirre:2016gsx}. 
Since the main parameters determining the flavor evolution are $\omega$ and $\mu$, with $\mu \gg \omega$, we can deduce that the fast  flavor instability  has a growth rate  proportional to $\mu$~\cite{Chakraborty:2016yeg,Izaguirre:2016gsx,Chakraborty:2016lct}. This explains why this  is expected to lead to  fast neutrino conversion. The linear stability analysis can be performed in the context of fast pairwise conversions through  the formalism described in Sec.~\ref{sec:LSA};  however, in the $\omega \rightarrow 0$ limit, a formally elegant dispersion relation in the flavor space can be obtained~\cite{Izaguirre:2016gsx}. 
\begin{marginnote}[]
\entry{Electron lepton number (ELN) crossings}{Crossings between the angular distributions of electron neutrinos and antineutrinos.}
\end{marginnote}

The favorable conditions triggering fast flavor instabilities [i.e., an ELN crossing or a non-negligible backward flux of (anti)neutrinos] are formally connected~\cite{Abbar:2017pkh}. In order to show this, let us  assume an axially symmetric and stationary configuration with the axis of symmetry along the radial direction and ignore the collision term; then, Eqs.~\ref{eq:EOM1} and \ref{eq:EOM2} become
\begin{eqnarray}
 \frac{\partial}{\partial r}\rho(\vec{x},\vec{p})= \frac{-\imath}{\cos\theta}[H(\vec{x},\vec{p}),\rho(\vec{x},\vec{p})]\ \mathrm{and}\ 
 \frac{\partial}{\partial r}\bar{\rho}(\vec{x},\vec{p})= \frac{-\imath}{\cos\theta}[\bar{H}(\vec{x},\vec{p}),\bar{\rho}(\vec{x},\vec{p})]\ ,
\end{eqnarray}
where  $\vec{x}$ is defined by $(r,\cos\theta)$ in polar coordinates, and  the term $\cos\theta$ in the denominator is due to the inner product of the neutrino velocity and the spatial gradient. 
If  $\cos\theta$ is absorbed in the Hamiltonian,  the condition required for fast neutrino conversions to occur is a crossing between $\rho_{ee}/\cos\theta$ and $\bar{\rho}_{ee}/\cos\theta$. Since, $\cos\theta$ can have negative values for a flux going radially inwards, this implies that spatial fast flavor instabilities exist  when a backward neutrino flux is present, even in the absence of ELN crossings. The magenta line in {\bf Fig.~\ref{Fig1}} shows the growth rate for the spatial instability (see Sec.~\ref{sec:LSA}). Due to the additional $\cos\theta$ factor in the denominator, the growth rate is substantially modified with respect to the blue line (temporal instability). 
In this sense, it is appropriate to consider {\it effective ELN crossings} in $(\rho_{ee}-\bar{\rho}_{ee})/\cos\theta$~\cite{Abbar:2017pkh}.

Since the evolution of fast neutrino conversions is dependent on the shape of the angular distributions, quantifying the entity of the flavor instability is not straightforward. However, a heuristic parameter, the so-called ELN parameter, has been proposed~\cite{Shalgar:2019qwg},
\begin{eqnarray}
\label{eq:xi}
\zeta = \frac{I_{1}I_{2}}{(I_{1}+I_{2})^{2}}
\label{zetadef}
\end{eqnarray}
with
\begin{eqnarray}
I_{1} = \int_{0}^{\pi} \Theta[\rho_{ee}(\theta) - \bar{\rho}_{ee}(\theta)]  d\cos\theta\ \mathrm{and}\ 
I_{2} = \int_{0}^{\pi} \Theta[\bar{\rho}_{ee}(\theta) - \rho_{ee}(\theta)]  d\cos\theta\ ,
\end{eqnarray}
where $\Theta$ is the Heavyside function. It can easily  be seen that $\zeta$ vanishes for no ELN crossings, however,  the growth rate may not be  proportional to $\zeta$. In the  steady-state configuration, an effective ELN parameter is obtained by replacing $\rho \rightarrow \rho/\cos\theta$.

We stress that, although the presence of effective ELN crossings typically gives rise to fast flavor instabilities, their existence is a necessary but not sufficient condition for the occurrence of  fast neutrino conversions; in particular, while each ELN crossing triggers an axially symmetry breaking instability, if  more than one crossing occur, it is not known whether  axially symmetric instabilities exist~\cite{Capozzi:2019lso,Stirner:2020}. A simple example, that can be verified analytically, is the one obtained  for the following initial conditions: $\rho_{ee}(\theta) = \textrm{const.}$ and $\bar{\rho}_{ee}(\theta) \propto \sin\theta$ for $\theta \in [0,\pi]$. This configuration can clearly give rise to two ELN crossings for the right normalization of $\bar{\rho}_{ee}$, but it   does not lead to fast flavor conversions.

The dispersion relation approach~\cite{Izaguirre:2016gsx} allows to classify fast flavor instabilities according to two categories~\cite{Capozzi:2017gqd}; an absolute instability,  is defined as  one which causes an exponential growth of the off-diagonal elements at the point of the initial perturbation, while a convective instability is such that  the off-diagonal elements decay at the point of the initial perturbation and the instability moves away faster than it spreads~\cite{Sturrock:1958zz,Capozzi:2017gqd,Yi:2019hrp,Capozzi:2019lso}.  
The existence of ELN crossings is expected to lead to an absolute instability~\cite{Capozzi:2017gqd}. 
Note, however, that the classification of an instability as convective or absolute is predicted  on the basis of the initial angular distributions;  this is an approximation as neutrino advection is not negligible  in a realistic framework. The impact of neutrino advection  on the classification of the nature of the  instability is largely unexplored~\cite{Capozzi:2019lso,Shalgar:2019rqe}.

The dispersion relation approach~\cite{Izaguirre:2016gsx}  has been extended to a three flavor framework in Refs.~\cite{Chakraborty:2019wxe,Capozzi:2020kge}, leading to an interesting phenomenology in the limit where the $\nu_\mu$  and $\nu_\tau$  fluxes are not identical. The dispersion relation has also been applied to scenarios involving  non-standard neutrino interactions~\cite{Dighe:2017sur}.

\begin{textbox}[h]
\subsection{Limitations of the linear stability analysis}
The linear stability analysis, and therefore the dispersion relation in the flavor space, predicts the existence of flavor instabilities on the basis of the initial setup of the neutrino ensemble at a given location $\vec{x}$ and time $t$; it is  a local analysis technique. The existence of flavor instabilities is  a necessary,  but not a sufficient, condition for  significant flavor conversions. The flavor conversion physics is  strongly affected by the non-linear regime of flavor conversions that is not captured by the linearized equations of motion.  
\end{textbox}

\section{PHENOMENOLOGY OF FAST PAIRWISE CONVERSIONS}\label{sec:pheno}

In this section, we  discuss the dependence of fast pairwise conversions on  neutrino energy and the eventual presence of  ELN crossings in the angular distributions. We then explore  the interplay between fast pairwise conversions, neutrino advection, and collisions.
In order to  explore the phenomenology of fast pairwise conversions,  we rely on the simple system  introduced in Eq.~\ref{eq:case1} (Case 1); 
in addition, we adopt $\mu = 10^3$~km$^{-1}$, $\Delta m^2 = + 2.5 \times 10^{-3}$~eV$^{2}$ (normal ordering), and $\theta_V = 10^{-6}$ to take into account the effective mixing suppression due to matter effects~\cite{EstebanPretel:2008ni}, unless otherwise specified.

\subsection{Dependence on neutrino energy}
Fast pairwise conversions have traditionally been explored by neglecting the vacuum term in the Hamiltonian, see e.g.~Ref.~\cite{Chakraborty:2016lct},  under the assumption that the vacuum frequency is negligible with respect to  the $\nu$--$\nu$ interaction strength. Reference~\cite{Airen:2018nvp} has extended the dispersion relation approach of Ref.~\cite{Izaguirre:2016gsx} to the most general case when slow and fast conversions can both occur. By focusing on the linear regime, the dispersion relation can exhibit fast modes for $\omega \neq 0$ and the growth rate of the instability can be $\mathcal{O}({\mu})$ (fast instability)  or $\mathcal{O}(\sqrt{|\omega \mu|})$ (slow instability).

\begin{figure}
\includegraphics[width=1.0\textwidth]{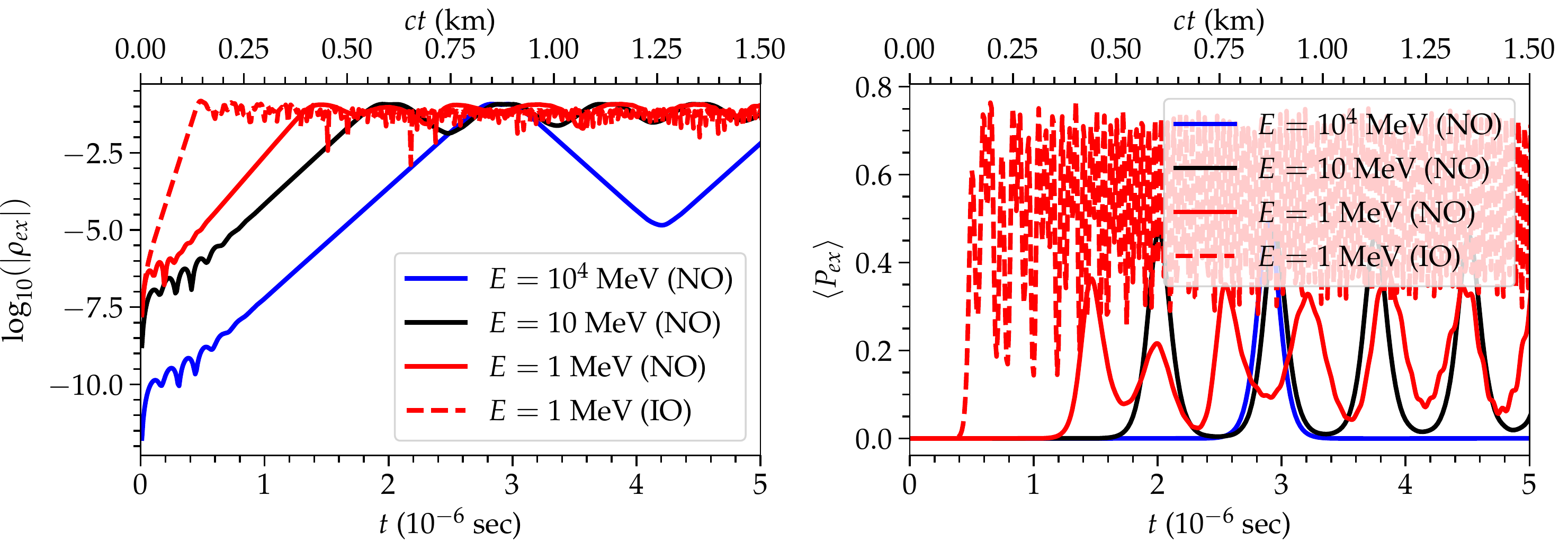}
\caption{Dependence of fast pairwise conversion on  neutrino energy and mass ordering. {\it Left:} Temporal evolution of the angle-integrated off-diagonal term of the density matrix for neutrino energies $E=1,10,10^4$~MeV (solid lines for normal mass ordering--NO, dashed lines for inverted mass ordering--IO). As $E$ decreases (the vacuum frequency $\omega$ increases),  an earlier onset of flavor conversion occurs. 
{\it Right:}  Temporal evolution of the angle-averaged transition probability. As $E$ decreases ($\omega$ increases),  the oscillation frequency increases in the non-linear regime.  The periodicity of flavor conversions is partially lost for large $\omega$.  
}
\label{fig:omega}
\end{figure}

\begin{figure}
\includegraphics[width=\textwidth]{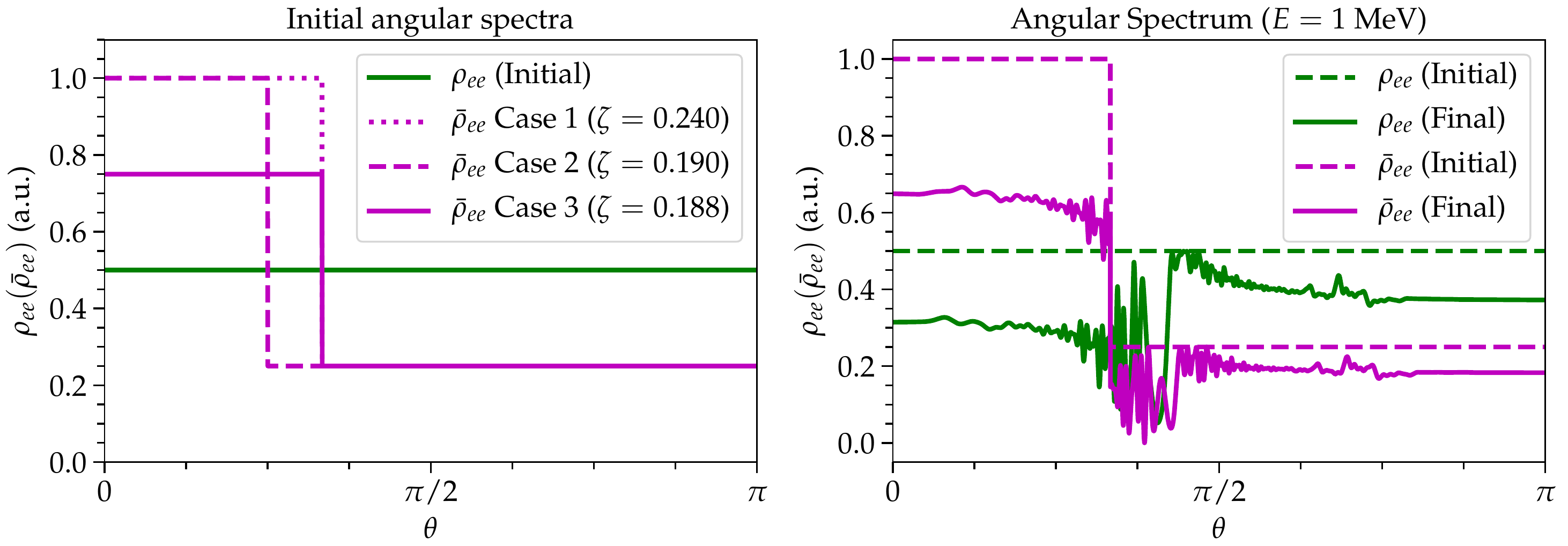}
\caption{ 
{\it Left:} Initial angular distributions for  Cases 1, 2 and 3  (see Sec.~\ref{sec:ELN} for details). The instability parameter $\zeta$ (Eq.~\ref{eq:xi}) is reported in the legend. 
{\it Right:} Angular distributions of $\nu_{e}$ and $\bar{\nu}_{e}$ for $E = 1$ MeV at $t=5\times10^{-6}$~s (solid lines) and $t=0$ (dashed lines). Flavor conversions develop in the proximity of the ELN crossing and spread in the neighbouring angular bins. 
}
\label{fig:ELN}
\end{figure}

Intriguingly, while the dependence of the linear growth rate on $\omega$ is not dramatic for neutrino energies typical  of neutrino dense environments, and therefore the approximation adopted in Refs.~\cite{Izaguirre:2016gsx,Chakraborty:2016lct} is good enough, the onset of flavor conversions is  affected by $\omega$~\cite{Dasgupta:2017oko,Johns:2019izj,Shalgar:2020xns}, as shown in the left panel of {\bf Fig.~\ref{fig:omega}} where the  angle-integrated off-diagonal term of the density matrix is plotted as a function of time for Case 1 and three different neutrino energies in normal ordering ($E=1, 10, 10^4$~MeV, where the latter case mimics the case of $\omega \rightarrow 0$ often adopted in the literature). 
The right panel of {\bf Fig.~\ref{fig:omega}} shows the angle-averaged transition probability:
\begin{eqnarray}
\label{eq:Pex}
\langle P_{ex} \rangle(t) =1 - \frac{\int \rho_{ee}(\theta,t)  d\cos\theta - \int \rho_{xx}(\theta,t=0)  d\cos\theta}
{\int \rho_{ee}(\theta,t=0)  d\cos\theta - \int \rho_{xx}(\theta,t=0)  d\cos\theta}\ ;
\end{eqnarray}
$\langle P_{ex} \rangle$  describes the average amount of flavor conversion despite the fact that the transition probability may be larger or smaller along any specific angular direction.
For $E=10, 10^4$~MeV, one can see a clear periodic trend in $\langle P_{ex} \rangle$~\cite{Dasgupta:2017oko,Johns:2019izj,Shalgar:2020xns}. In addition, an earlier onset of flavor conversions and an increase of  the oscillation frequency occur as $\omega$ increases~\cite{Dasgupta:2017oko,Shalgar:2020xns}. This is due to the fact that the system is driven by two characteristic frequencies $\omega$ and $\mu$ and, as $\omega$ increases, the onset of the non-linear regime occurs earlier.  
Hence the pendulum analogy, introduced to explain $\nu$--$\nu$ interactions in the case of slow conversions~\cite{Hannestad:2006nj}, does not hold for fast pairwise conversions unless $\omega \rightarrow 0$~\cite{Dasgupta:2017oko,Johns:2019izj,Shalgar:2020xns}.
For $E=1$~MeV, the oscillation frequency is higher and the bipolar behavior is partially lost due to the overlap of two different frequencies dominating the precession~\cite{Shalgar:2020xns}. 
For Case 1 with $E=1$~MeV, {\bf Fig.~\ref{fig:ELN}}  displays the angular distributions of $\nu_e$ and $\bar\nu_e$ before (left panel) and after (right panel) fast pairwise conversion. One can clearly see that the flavor instability arises in the proximity of the ELN angular crossing and it spreads through the neighbouring angular bins.


Contrarily to slow neutrino self-interactions, a peculiar feature of fast pairwise conversions is that they are not strongly affected by the energy distribution of (anti)neutrinos, and the typical average vacuum frequency describes well the behavior of the system~\cite{Shalgar:2020xns}. However, as shown by the comparison between the solid and the dashed lines for $E=1$~MeV in the left panel of {\bf Fig.~\ref{fig:omega}} (obtained for normal and inverted mass ordering, respectively), the growth rate is steeper in inverted ordering and the onset of flavor conversions is reached earlier  for the setup we consider,  in analogy to the slow conversion case~\cite{Duan:2006an,Hannestad:2006nj}.

\subsection{Dependence on electron lepton number crossing}\label{sec:ELN}

The occurrence of ELN crossings is deemed to be one of the crucial factors triggering fast pairwise conversion~\cite{Izaguirre:2016gsx}.  The dependence of flavor conversion   on ELN crossing has been explored in Refs.~\cite{Yi:2019hrp,Martin:2019gxb} in the linear regime. By relying on the development and evolution of the branches of the  dispersion relation,  as the ELN distribution changes continuously,  a theory of the critical points of the dispersion relation  has been elaborated  to classify the  instability type~\cite{Yi:2019hrp}. 
However, for flavor conversions to grow, the ELN crossings have to be self-sustained in time~\cite{Shalgar:2019qwg}. For example, neutrino advection  can smear the ELN crossings hindering the development of fast pairwise conversions.

In order to explore how the flavor conversion physics is affected by the ELN crossings in the non-linear regime, we explore other two configurations (Cases 2 and 3) in addition to Case 1 introduced in Eq.~\ref{eq:case1}. Cases 2 and 3 are defined  by keeping $\rho_{ee}(t=0)$  the same as in Case 1, $\rho_{xx}(t=0)=\bar\rho_{xx}(t=0)=0$, and   
\begin{eqnarray}
\bar{\rho}_{ee,\ \mathrm{Case 2}}(t=0) &=& 
\begin{cases}
2 \rho_{ee}(t=0) & \theta \in [0,\pi/4]  \\
0.5\rho_{ee}(t=0) & \theta \in [\pi/4,\pi]\ \end{cases}\ ,   \nonumber\\ 
\bar{\rho}_{ee,\ \mathrm{Case 3}}(t=0) &=& 
\begin{cases}
1.5 \rho_{ee}(t=0) & \theta \in [0,\pi/3]  \\
0.5\rho_{ee}(t=0) & \theta \in [\pi/3,\pi]\ \end{cases} \ .
\end{eqnarray}
Cases 1, 2, and 3 are shown in the left panel of {\bf Fig.~\ref{fig:ELN}}; note that  the width (height) of the top-hat distribution of $\bar\nu_e$ is different  in Case 2 (Case 3) with respect to Case 1. Hence, the three Cases  display different ELN crossings, as indicated by the  instability parameter in the plot legend (see Eq.~\ref{eq:xi}). 

\begin{figure}
\includegraphics[width=\textwidth]{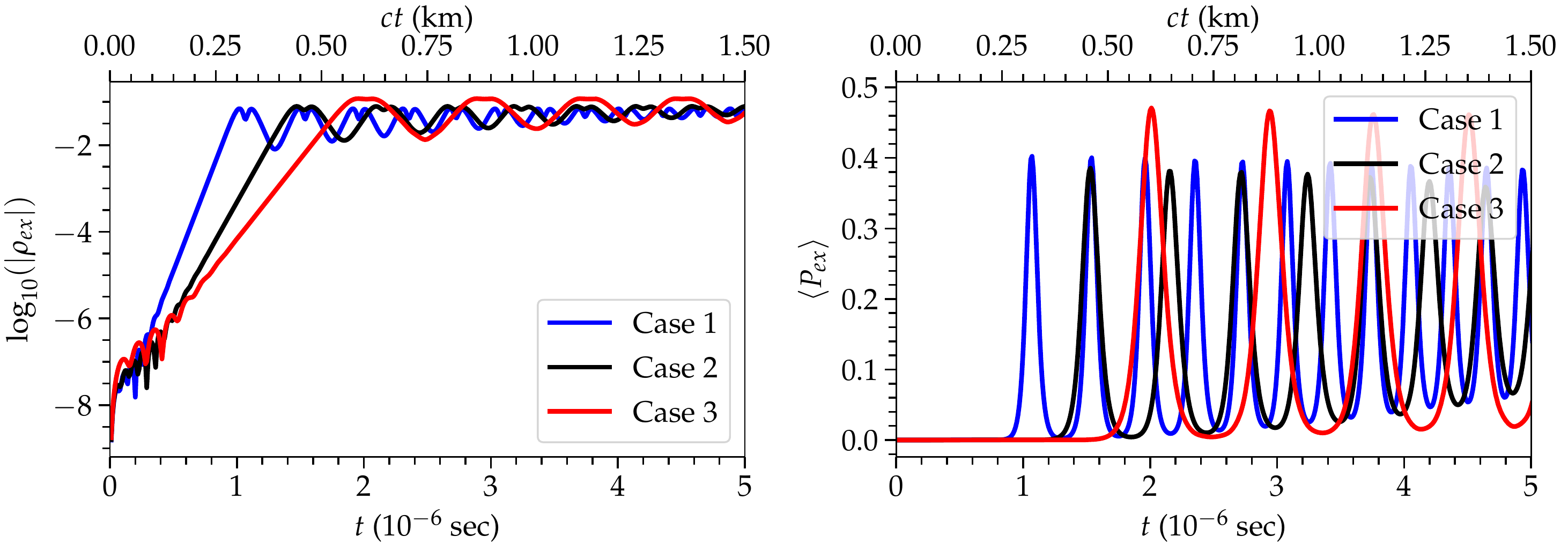}
\caption{Dependence of fast pairwise conversion on  electron lepton number crossing.
{\it Left:} Temporal evolution of the angle-integrated off-diagonal term of the density matrix for  Cases 1, 2, and 3 (see Sec.~\ref{sec:ELN} for details). The onset of the non-linear regime occurs earlier for larger $\zeta$.  {\it Right:} Temporal evolution of the angle-averaged transition probability. The oscillation amplitude does not grow monotonically with $\zeta$.  
}
\label{fig:angdist}
\end{figure}

The left (right) panel of {\bf Fig.~\ref{fig:angdist}} shows the temporal evolution of the angle-averaged off-diagonal term of the density matrix (survival probability) for Cases 1, 2, and 3. One can see that, for larger $\zeta$, the onset of flavor conversions occurs earlier.  Note, however, that the transition probability has the largest oscillation amplitude for the smallest $\zeta$ parameter (Case 3) and it is not a monotonic function of $\zeta$~\cite{Shalgar:2019qwg}. The behavior of Case 3 highlights the limitations intrinsic to the predictive power of  $\zeta$; in fact, in order to fully predict the flavor conversion outcome, one should  take into account the slope in addition to the width of the ELN crossing.

\subsection{Dependence on neutrino advection and collisions}\label{sec:collisions}
The angular distributions of (anti)neutrinos play a major role in determining the development of  fast neutrino conversions, but the origin of the angular distributions is a largely unexplored topic in the context of flavor conversions, see e.g.~Refs.~\cite{Capozzi:2018clo,Shalgar:2019kzy,Shalgar:2020wcx} as examples of approximate implementations. 
The collision term in the equations of motion is used to incorporate  absorption, emission, and  momentum changing scatterings of neutrinos. The spatial distribution of these terms in conjunction with the advective term ($\vec{v}\cdot\vec{\nabla}_x$) determines the angular distribution of neutrinos at any given location. A flavor dependent collision term has the effect of reducing the coherence between the flavor eigenstates, while a flavor independent collision term will transport all angular modes.

In the deep interior of compact objects, the angular distributions of all neutrino flavors are isotropic due to the high collisional rate.  As the matter density decreases,  neutrinos approach the free-streaming regime and their angular distribution becomes forward peaked~\cite{Brandt:2010xa,Tamborra:2017ubu}. The intermediate region, i.e.~before decoupling is complete,  is of significance to fast conversions as   flavor evolution, advection, and collisions may be  at play  simultaneously~\cite{Johns:2019izj,Capozzi:2018clo,Tamborra:2017ubu}. 
In the earlier literature on $\nu$--$\nu$ interactions, neutrinos were assumed to decouple at a well defined radius~\cite{Duan:2006an}. This is a reasonable assumption in the case of slow $\nu$--$\nu$ interaction as  neutrino flavor conversions occur away from the decoupling region (see, e.g., {\bf Fig.~\ref{fig:SNsketch}}). For fast conversion, however, this assumption needs a renewed scrutiny as the distinction between the region of neutrino trapping (isotropic angular distributions) and free-streaming (forward-peaked angular distributions) is gradual and fast flavor conversions may occur in between. 

The $\bar\nu_e$ cross-section  is typically smaller than the $\nu_e$ one; hence, $\bar\nu_e$'s start decoupling earlier than $\nu_e$'s. This can result in ELN crossings as discussed in Ref.~\cite{Shalgar:2019kzy}. Reference~\cite{Shalgar:2019kzy}, however, did not consider  the occurrence of ELN crossings (and therefore flavor conversions) eventually due to  large-scale asymmetries~\cite{Tamborra:2014aua} or turbulent matter density fluctuations~\cite{Abbar:2020ror}. 
The occurrence of ELN crossings or lack thereof depends on the ratio of electron neutrino and antineutrino number densities~\cite{Shalgar:2019kzy}. A large difference between $n_{\nu_e}$ and $n_{\bar\nu_e}$  implies that no ELN crossing can occur. However, if ELN crossings exist, flavor conversions may dynamically modify the angular distributions, and in turn the ELN crossings. Importantly, the ELN crossings are also dynamically  affected  by collisions~\cite{Shalgar:2019kzy}.

\begin{figure}
\includegraphics[width=\textwidth]{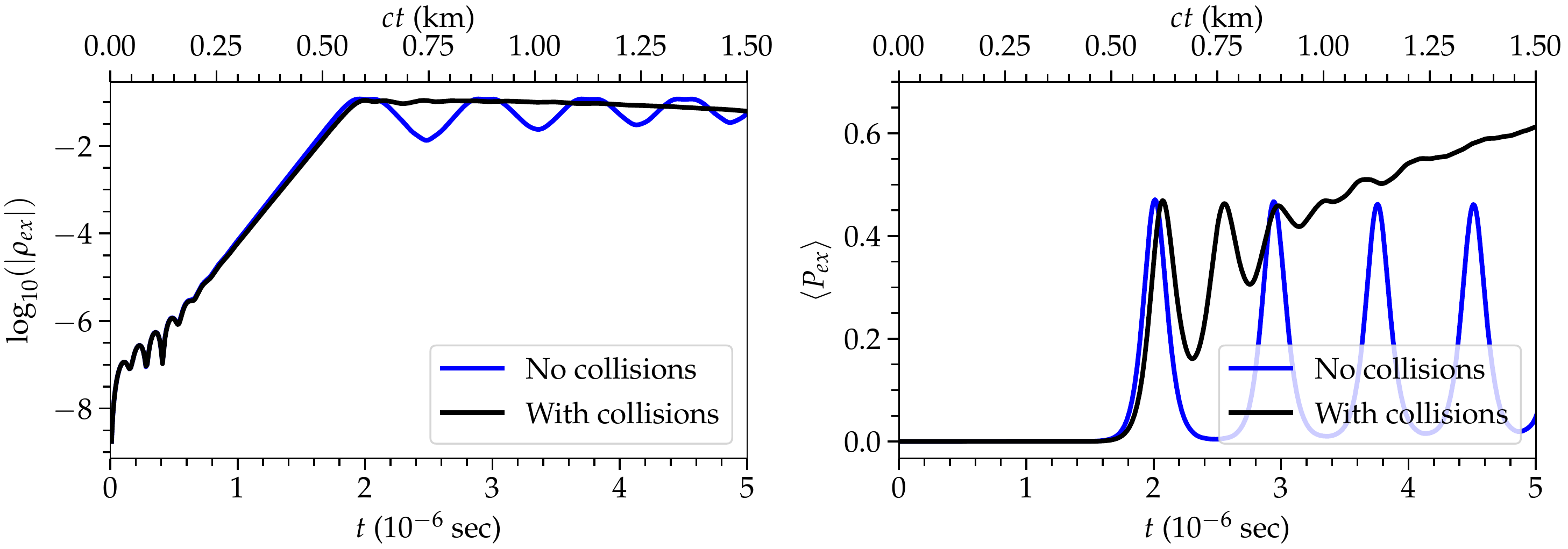}
\caption{Dependence of fast pairwise conversion on collisions. {\it Left:} Temporal evolution of the angle-integrated off-diagonal term of the density matrix with (black line) and without (blue line) collisions (see Sec.~\ref{sec:collisions} for details). The collision strength is so weak that the onset of the non-linear regime is not affected.   {\it Right:} Temporal evolution of the angle-averaged transition probability. The flavor conversion probability is  enhanced in the presence of collisions. }
\label{fig:collisions}
\end{figure}

The effect of collisions on fast neutrino conversions is largely unexplored despite being probably relevant in triggering flavor conversions~\cite{Capozzi:2018clo} and determining the final flavor outcome~\cite{Shalgar:2020wcx}. For simplicity, we assume a flavor independent collision term, which is number conserving and energy independent, and assume that the collision term for neutrinos is twice  the one for antineutrinos and is equal to the inverse of the neutrino mean free path for all angular bins ($\mathcal{C}=  \bar{\mathcal{C}}/2= 1$~km$^{-1}$) for  Case 1 with $E=10$~MeV. As shown in  {\bf Fig.~\ref{fig:collisions}}, the collision term can significantly enhance the flavor conversion probability because of its dynamical effect on  the angular distributions~\cite{Shalgar:2020wcx}. 
The results shown in {\bf Fig.~\ref{fig:collisions}} ignore the advective term in the equations of motion. Advection  could mix different angular modes as a function of time. Although obtained within a simplified framework,  {\bf Fig.~\ref{fig:collisions}} suggests a possible non-negligible   interplay of fast flavor conversions with collisions and  advection; this may imply the necessity to tackle the flavor evolution as a time-dependent boundary value problem. 

\begin{textbox}[h!]
\subsection{Numerical artifacts in the evolution of neutrino flavor}
It is well known that multi-angle calculations of $\nu$--$\nu$ interactions may be plagued by spurious instabilities~\cite{Sarikas:2012ad,Morinaga:2018aug}; this problem could also arise in the case of fast pairwise conversions. In addition, it has been shown that fast  conversions may enhance a cascade of flavor field power from large angular scales to small scales, hastening relaxation~\cite{Johns:2020qsk,Bhattacharyya:2020jpj}.  The momentum-space cascade can then induce numerical errors that may propagate back to larger scales, with possible major consequences on the isotropic moment. This problem has not yet been studied in the presence of collisions and it is therefore not known whether collisions could partially cure the back-reaction triggered by the  instabilities that develop on small scales. 
\end{textbox}

\section{FAST PAIRWISE CONVERSIONS IN CORE-COLLAPSE SUPERNOVAE}\label{sec:SN}
In this section, after a brief overview on  core-collapse supernova physics, the occurrence of favorable conditions for the development of fast pairwise conversions in core-collapse supernovae is presented, together with their possible implications.

\subsection{Core-collapse supernovae}
Core-collapse supernovae represent the final stage of the life of stars with mass of at least $8\ M_\odot$~\cite{Vitagliano:2019yzm,Janka:2006fh,Janka:2017vcp,Burrows:2012ew,Burrows:2020qrp}. A supernova is effectively a blackbody source of (anti)neutrinos of all flavors; when the stellar core collapses, neutrinos carry $99\%$ of the gravitational binding energy ($E_b \sim 3 \times 10^{53}$~erg). 
Our current understanding of the stellar collapse is based on the delayed neutrino-driven explosion mechanism~\cite{Bethe:1984ux}. 
 The massive star has an onion-like structure with an iron core and lighter elements in the outer shells. The iron core undergoes a homologous collapse until it reaches nuclear density of $\mathcal{O}(3\times 10^{14})$~g~cm$^{-3}$, at which point the sudden halt of the collapse results in a shock-wave~\cite{shapiro2008black}.
 The shock-wave propagates radially outwards, but it stalls  as it loses energy by photo-dissociating iron group nuclei. Neutrinos revive the shock by  depositing energy, and a successful explosion takes place.

The typical neutrino signal lasts for $\mathcal{O}(10)$~s, and the non-electron neutrinos and antineutrinos have very similar energy distributions because of the  typical energies involved, although some muons could be present~\cite{Bollig:2017lki}.
At first, the prompt neutronization burst of  $\nu_e$'s occurs when the shock wave breaks through the iron shell. 
The accretion phase follows, during which $\nu_e$ and $\bar\nu_e$ are emitted  with similar energy luminosities and different average energies. The main production and interaction channel goes through beta reactions involving the electron flavors, while the non-electron flavors are produced in pairs in the inner layers and have lower energy luminosities. Soon after the explosion, the newly formed neutron star cools and deleptonizes, and  the neutrino fluxes of all flavors become similar to each other. 
We refer the reader to, e.g.,~Refs.~\cite{Vitagliano:2019yzm,Mirizzi:2015eza,Janka:2006fh,Burrows:2020qrp} for detailed overviews.

Hydrodynamical simulations of the core collapse have  approached the 3D frontier~\cite{Janka:2016fox,Burrows:2019xff}. Given the challenges involved in the modeling of the neutrino flavor conversion physics,  the neutrino transport equations in hydrodynamical simulations do not include  flavor conversions~\cite{Mezzacappa:2020oyq}. Instead, flavor conversions are  investigated in a post-processing phase. This is justified because, according to the classic picture, MSW  conversions and $\nu$--$\nu$ interactions in the slow regime should occur beyond the shock radius  in a spherically symmetric supernova model, as shown in \textbf{Fig.~\ref{fig:SNsketch}}. As such,   flavor conversions would only be relevant for detection purposes and for the nucleosynthesis in the neutrino-driven wind~\cite{Mirizzi:2015eza}. 
 However, this simplified picture neglects the occurrence of  large-scale  asymmetries~\cite{Janka:2016fox,Vartanyan:2019ssu} or effective ELN crossings; fast  flavor conversions  could then potentially occur  in the proximity of the decoupling region, during the supernova accretion phase, possibly affecting  the supernova physics itself~\cite{Sawyer:2015dsa,Izaguirre:2016gsx}.

\subsection{Fast pairwise conversions in supernovae}
Because of the  potential implications of fast pairwise conversions on the supernova physics, various groups have looked for the occurrence of ELN crossings in hydrodynamical simulations of core-collapse supernovae. The first attempt was carried out in Ref.~\cite{Tamborra:2017ubu} that looked for ELN crossings in a set of spherically symmetric (1D) hydrodynamical simulations; no evidence of ELN crossings was found. 
Multi-D simulations are naturally more prone to develop large-scale asymmetries in the ELN emission, e.g.~in the presence of LESA~\cite{Tamborra:2014aua}. However, except than for some instances, e.g.~Ref.~\cite{Nagakura:2017mnp}, most of the available multi-D simulations, e.g.~\cite{Tamborra:2014aua,Bruenn:2014qea,OConnor:2015rwy,Nagakura:2017mnp,Richers:2017awc,Vartanyan:2018xcd,Just:2018djz,Cabezon:2018lpr,Pan:2018vkx,Glas:2018vcs,Glas:2018oyz}, only track the ``moments,'' i.e.~the energy-dependent angular integrals of the neutrino phase space distributions, because it is computationally too demanding to store and self-consistently compute the  fully angle-dependent distributions as  functions of time. Therefore, to diagnose the possible presence of ELN crossings
 (which, however, does not automatically imply the existence of flavor instabilities), Refs.~\cite{Dasgupta:2018ulw,Abbar:2020fcl} have proposed alternative methods. For example,     a Fourier mode of the flavor instability, the ``zero mode'' which  has a growth rate depending on the ELN angular moments up to the second order, was identified in~\cite{Dasgupta:2018ulw}.  The growth rate of this mode approximates the one of flavor conversions~\cite{Dasgupta:2018ulw}. Along the same lines, Ref.~\cite{Abbar:2020fcl} recently proposed a method based on higher angular moments.

As discussed in Ref.~\cite{Shalgar:2019kzy}, ELN crossings are expected to occur when the asymmetry parameter $\gamma = n_{\bar\nu_e}/n_{\nu_e} \simeq 1$. References~\cite{Abbar:2018shq, Azari:2019jvr}  found ELN crossings deep in the proto-neutron star region for a number of isolated points where $\gamma \simeq 1$. They linked the occurrence  of ELN crossings with locations where the $\nu_e$ chemical potential  $\mu_{\nu_e} \simeq 0$ and the electron fraction is relatively low. In addition, Ref.~\cite{DelfanAzari:2019tez} suggested that the appearance of light nuclei (mostly $\alpha$ particles) may support the development of ELN crossings by enhancing the chemical potential difference between protons and neutrons.

A similar analysis has been carried out in Ref.~\cite{Glas:2019ijo} which searched for ELN crossings by adopting the method proposed in Ref.~\cite{Dasgupta:2018ulw}. Favorable conditions for the development of  fast flavor instabilities  were found deep in the convective layer of the proto-neutron star;  the decline of the electron fraction and the increase of density and temperature drive the  electron-neutrino chemical potential to negative values and hence an excess of $\bar\nu_e$ over $\nu_e$ forms. The findings of Ref.~\cite{Glas:2019ijo} confirm the overall conclusions of Refs.~\cite{DelfanAzari:2019tez,Abbar:2019zoq}. However, Ref.~\cite{Glas:2019ijo} points out that the  spatial locations of the ELN crossings develop in time within the boundary layers of large-scale and long-lasting volumes  where the $\bar\nu_e$ density exceeds the $\nu_e$ one. 
\begin{marginnote}[]
\entry{Lepton-Emission Self-Sustained Asymmetry (LESA)}{Neutrino-driven hydrodynamical instability responsible for the asymmetric emission of electron neutrinos and antineutrinos.}
\entry{$\alpha$ particle}{Helium-4 nucleus (made of two protons and two nucleons).}
\entry{Proto-neutron star}{Hot compact remnant formed in the early supernova phase that later cools and deleptonizes, leading to a neutron star.}
\entry{Electron lepton fraction}{Ratio between the effective number density of electrons and the one of baryons.}
\end{marginnote}

The ELN crossings may also occur in the post-shock flows~\cite{Nagakura:2019sig}; in  the proximity of the proto-neutron star, 
$\bar\nu_e$  dominates over $\nu_e$ in the forward direction, while the converse is true at larger radii where $\nu_e$ dominates over $\bar\nu_e$ in the forward direction because of the residual coherent neutrino-nucleus scattering~\cite{Morinaga:2019wsv}. 
 The small ELN crossings generated at large radii seem to lead to flavor instabilities according to the stability analysis. This would not have major consequences on the supernova dynamics, but it may still affect the nucleosynthesis in the neutrino-driven ejecta.
 
 The  consequences of the occurrence of fast pairwise conversions on the supernova neutrino-driven wind   nucleosynthesis  have been explored under the extreme assumption that fast pairwise conversions may lead to flavor equilibration~\cite{Xiong:2020ntn}. Flavor equilibration  may  significantly favor  proton-rich conditions with an enhancement of the total mass loss by a factor $\mathcal{O}(1.5)$. This would have important implications for abundances in metal-poor stars and Galactic chemical evolution.
 
 It is worth noticing that, if fast pairwise conversions should occur in the proto-neutron star layer and  its surroundings, they would impact the neutrino spectra formation and the supernova dynamical evolution. Moreover, neutrino advection could smear the ELN crossings~\cite{Shalgar:2019qwg}, unless they are self-sustained, and therefore hinder the development of fast pairwise conversions. This section should then be considered as indicative of potentially interesting effects of pairwise conversions on the supernova physics, but only a fully self-consistent solution of the neutrino transport could shed light on the role of neutrino conversions in the stellar core and in the supernova nucleosynthesis.

\section{FAST PAIRWISE CONVERSIONS IN NEURON STAR MERGERS}\label{sec:NSM}
In this section, after a brief overview on the neutron star merger physics, the role of fast pairwise conversions is outlined together with the implications for the nucleosynthesis of the heavy elements.

\subsection{Compact binary merger remnants}

A compact binary merger originates by the coalescence of two neutron stars or a neutron star and a black hole. The central remnant could be a massive neutron star or a black hole. Because of the merging,  an accretion disk forms surrounding the central remnant. Neutrinos are copiously produced during the coalescence, and the neutrino energy luminosity may reach up to $10^{54}$~erg/s within $\mathcal{O}(100)$~ms~\cite{Ruffert:1996by,Foucart:2015vpa}. 

Due to the low local merger rate, the probability of detecting thermal neutrinos from neutron star mergers is negligible~\cite{Abe:2018mic}. However, neutrinos may play an important indirect role. The neutron richness of the ejecta may be affected by neutrino absorption on  matter dynamically ejected during the  merger; also,  the $r$-process nucleosynthesis and the related kilonova lightcurve may be impacted by the neutrino field~\cite{1974ApJ192L.145L,Eichler:1989ve,Li:1998bw,Kulkarni:2005jw,Metzger:2010sy}.
In addition, neutrinos play an important role in the merger cooling, and  neutrino pair annihilation may aid  the short gamma-ray burst harbored by the merger~\cite{Eichler:1989ve,Wanajo:2014wha,Perego:2014fma,Fernandez:2015use,Sekiguchi:2015dma,Radice:2016dwd,Miller:2019dpt,Woosley:1993wj,Ruffert:1998qg,2011MNRAS.410.2302Z,Just:2015dba,Foucart:2020qjb}. Multi-messenger observations~\cite{TheLIGOScientific:2017qsa,Monitor:2017mdv,GBM:2017lvd} of the gravitational wave event GW 170817 have confirmed the theory according to which  compact binary mergers are among the main sites where the elements heavier than iron form through  $r$-process; however, many unknowns  remain.
\begin{marginnote}[]
\entry{$r$-process}{Rapid neutron capture process invoked to explain the origin of the elements heavier than iron.}
\end{marginnote}

\begin{figure}[t]
\includegraphics[width=3.5in]{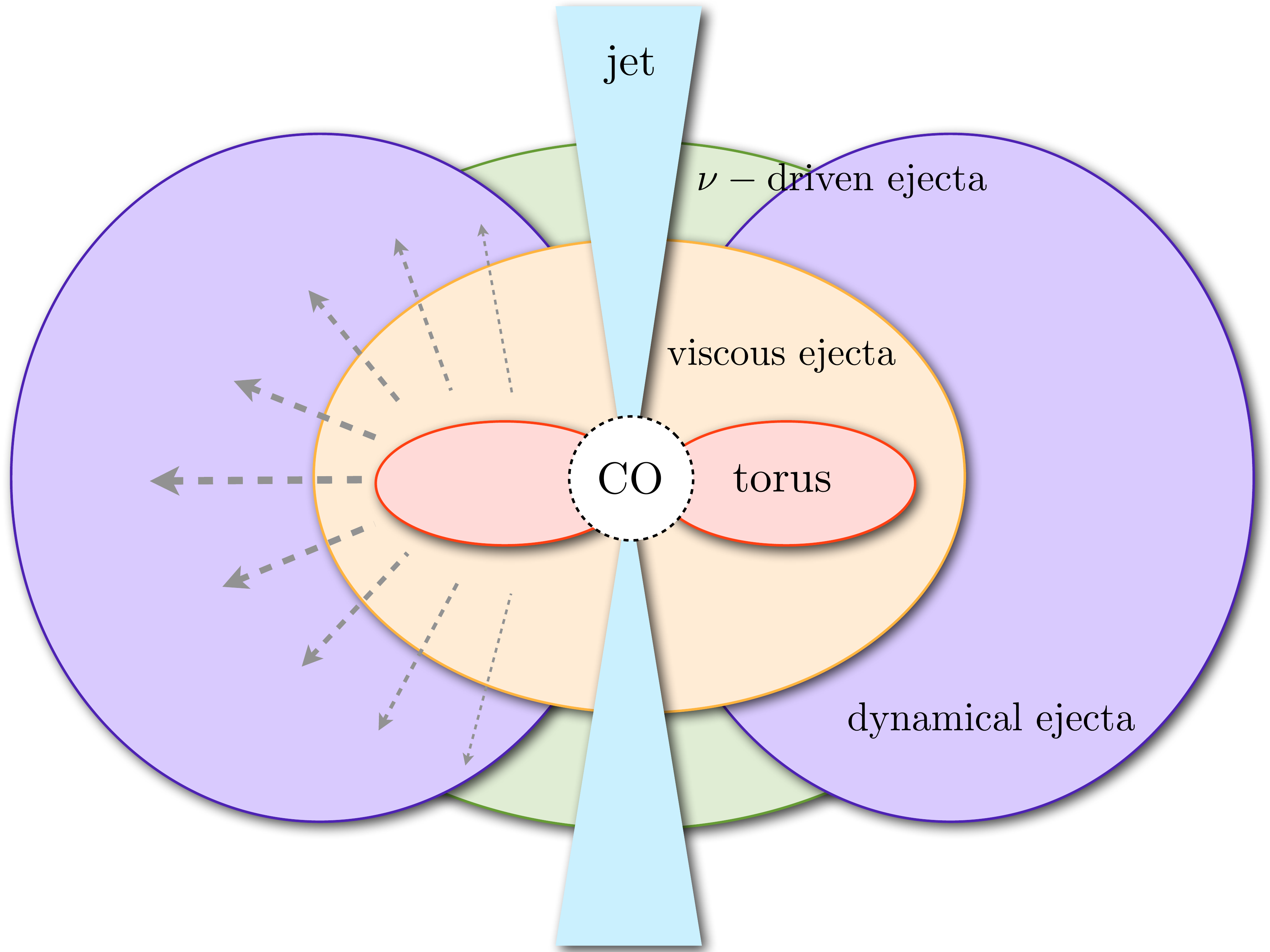}
\caption{Schematic representation of the torus remnant and its ejecta. The central compact object (CO) can be a hypermassive neutron star or a black hole; the picture is qualitatively similar in both cases, with a more massive neutrino-driven wind expected for the hypermassive neutron star remnant.  The dynamical ejecta (violet-shaded area) are the earliest matter
    outflows, followed by the neutrino-driven  (green-shaded area) and the 
    viscously driven  (orange-shaded area) ejecta.   The neutrino-driven wind may dominate the ejecta in the surroundings of the polar axis. Figure adapted from Ref.~\cite{Wu:2017drk}.}
\label{fig:mergers}
\end{figure}

Once the accretion disk forms, up to $20\%$ of the initial disk mass can be ejected. 
The dynamical ejecta are the earliest matter outflows~\cite{Foucart:2014nda,Radice:2016dwd,Sekiguchi:2015dma}; they originate from the outermost layers  which are  unbounded by means of tidal torques. 
For the first few hundred milliseconds,  a neutrino-driven wind is emitted from  the hot  inner disk~\cite{Perego:2014fma,Just:2014fka}. Within a few seconds, viscously-driven ejecta are then emitted~\cite{Just:2014fka,Fernandez:2013tya,Lippuner:2017bfm,Fujibayashi:2017puw}.
The neutrino-driven wind can  be  relevant in the surroundings of the  polar region, where it dominates the  ejecta in a cone centered around the polar axis with half-opening angle of  $10$--$40^\circ$~\cite{Wu:2017drk}. This picture is schematically represented in 
{\bf Fig.~\ref{fig:mergers}} and it holds independently on the nature of the central compact object, although   a more massive neutrino-driven wind is expected in the case of a hypermassive neutron star remnant.

The  evolution of the accretion disk can be divided in three stages as for the neutrino emission properties~\cite{Just:2014fka}. At first,  the environment is dense enough to be optically thick to neutrinos; neutrinos are  trapped and advected in the flow and  cooling is  inefficient.
This phase is followed by a period of neutrino-dominated accretion, when the mass of the torus decreases and the density drops;  neutrinos radiate most of the gravitational energy converted in internal energy. As mass, temperature, and density decrease, the neutrino production rate decreases until  neutrino cooling eventually becomes inefficient. 
When the remnant compact object consists of a  massive neutron star instead than a black hole, the neutrino energy luminosity  reaches a plateau, instead than decreasing in time~\cite{Perego:2014fma,Ardevol-Pulpillo:2018btx,George:2020veu}. 

The computational requirements for running three-dimensional, general-relativistic magnetohydrodynamical simulations of compact binary mergers with detailed neutrino transport are not yet available. Hence, the exploration of the role of neutrinos in merger remnants is  extremely preliminary. A generic feature of all simulations is the protonization of the merger remnant, leading to an excess of $\bar\nu_e$ over $\nu_e$. Since the neutrino density  in compact binary mergers is comparable to the one of core-collapse supernovae,  $\nu$--$\nu$ interactions should not be  negligible. Because of the disk protonization, a cancelation between the $\nu$--$\nu$ interaction strength and the matter one    may occur in the neutrino equations of motion. This  phenomenon is known as matter-neutrino resonance and  may affect the final flavor configuration~\cite{Malkus:2014iqa,Malkus:2012ts,Wu:2015fga,Tian:2017xbr,Vlasenko:2018irq,Shalgar:2017pzd}.

\subsection{Fast pairwise conversions in compact binary mergers}

In neutron star mergers, like in supernovae, the $\bar\nu_e$ decoupling occurs at  radii smaller than  the $\nu_e$ one. However, the overall flux of $\bar\nu_e$ is larger than the one of $\nu_e$. As a consequence of the disk protonization and of the toroidal geometry, ELN crossings occur anywhere above the disk remnant~\cite{Wu:2017qpc}.  
By relying on the linear stability analysis, one finds that fast flavor instabilities should occur everywhere above the merger remnant~\cite{Wu:2017qpc,Wu:2017drk}. In particular, while temporal flavor instabilities are expected in an extended region above the remnant because of the ubiquitous appearance of ELN crossings, spatial instabilities take place in smaller spatial regions (overlapping with the ones of temporal instabilities) but with a larger growth rate.
In the case of a black hole remnant, the region where flavor instabilities occur tends to shrink after the first $\mathcal{O}(10)$~ms~\cite{Wu:2017drk}. While the region where the instabilities occur remains stable for a longer time interval for a massive neutron star remnant~\cite{George:2020veu}. 

The occurrence of fast flavor instabilities~\cite{Wu:2017drk,Wu:2017qpc}, however, does not imply that flavor equilibration can be  achieved. For example, by relying on a simplified toy model, Ref.~\cite{Padilla-Gay:2020uxa}  explored the non-linear regime of fast pairwise conversions above the merger remnant disk and found  that negligible mixing is achieved  ($<1\%$) despite the  large growth rate of the fast flavor instability. This result should not be considered as a firm conclusion since the collisional term in the equations of motion was neglected in Ref.~\cite{Padilla-Gay:2020uxa} and this could enhance the flavor conversion probability~\cite{Shalgar:2020wcx}; however, it suggests that we are far from having a clear picture of fast pairwise conversions in compact binary mergers.

If we assume that flavor equilibration is achieved due to fast pairwise conversion above the merger remnant disk,  this could have consequences on the $r$-process nucleosynthesis.
Reference~\cite{Wu:2017drk} has explored the  impact of flavor equilibration on a black hole binary merger remnant and found that the fraction of lanthanides produced in the neutrino-driven wind could be boosted by a factor of $\mathcal{O}(10^3)$ with possible implications on the  kilonova emission. 
On the other hand, in the case of a massive neutron star compact merger remnant, Ref.~\cite{George:2020veu} has found that flavor equipartition may induce variations in the polar ejecta by enhancing the iron peak abundances and reducing the  first peak abundances, but the fraction of emitted lanthanides is negligibly affected. The different outcome in the two configurations is mostly due to the different dynamical evolution of the two remnant models, being the trajectories exposed for longer time to the neutrino wind in the  black hole merger remnant case, despite the flavor conversion outcome being very similar.

\section{FAST PAIRWISE CONVERSIONS IN THE EARLY UNIVERSE}\label{sec:EU}

The early universe is rich in neutrinos~\cite{Vitagliano:2019yzm,lesgourgues2013neutrino,Zyla:2020zbs}. The relic background of neutrinos from the early universe has not been detected yet, but we have indirect evidence from the Big Bang Nucleosynthesis, the Cosmic Microave Background anisotropies, and the formation of cosmic structures. It is important to grasp the neutrino flavor conversion physics in the early universe as this could have major implications on cosmological observables such as the effective number of radiation species ($N_{\mathrm{eff}}$),  the eventual existence of dark radiation, and indirectly  the Hubble parameter ($H_0$) as well as  the sum of the neutrino masses. 

 In the absence of physics beyond the Standard Model, the lepton asymmetry is expected to be of the same order as the baryon asymmetry, making $\nu$--$\nu$ interactions  negligible. However, it is possible that electron-positron annihilation favors an excess of $\nu_e$ and $\bar\nu_e$ over the other flavors with  potential implications on  neutrino mixing~\cite{Akita:2020szl}. According to the standard picture, the  excess of $\nu_e$'s and $\bar\nu_e$'s is small enough to have a negligible effect on the flavor mixing; this finding has also been confirmed by  relaxing the assumption of small lepton asymmetry and assuming  homogeneity and isotropy~\cite{Lunardini:2001pb,Dolgov:2002ab, Abazajian:2002qx, Wong:2002fa}. However, it is worth noticing that current limits on the lepton asymmetry assume that we know how to calculate  neutrino flavor conversions in the early universe and how flavor conversion, in turn,  affects the primordial abundances.
In the context of slow neutrino self-interactions, by adopting the linear stability analysis, it has been found that anisotropic and inhomogeneous modes can lead to the development of instabilities despite the existence of approximately isotropic and homogeneous  initial conditions~\cite{Cirigliano:2017hmk}. This suggests that synchronized flavor conversion~\cite{Sigl:1992fn,Dolgov:2002ab,Pastor:2001iu,Froustey:2020mcq} may just be a part of a much bigger picture~\cite{Johns:2016enc,Cirigliano:2017hmk}, and doubts are casted on the lepton asymmetry limits~\cite{Hansen:2020vgm}.

Numerical work on inhomogeneous and anisotropic neutrino flavor modes in the non-linear regime in the presence of a self-consistent treatment of collisions has not been attempted yet. The vacuum frequency scales as the inverse of the temperature, while the collision term depends on the fifth power of temperature~\cite{Bell:1998ds}; this leads to an enormous range of scales over which numerical simulations have to be performed, making the problem technically challenging to solve. 
However, the presence of  a transient localized flavor instability along with collisions and advection could trigger fast flavor conversions. 
The presence of  inhomogeneous modes in the context of fast flavor conversions could affect the entropy evolution in the early universe in yet unknown ways. 
An analysis of this sort  may bring new insights on the flavor conversion physics in the early universe and related cosmological observables, if fast pairwise conversions should induce large mixing.

\begin{summary}[SUMMARY POINTS] 
\begin{enumerate}
\item  Neutrino flavor conversion may play a crucial role in dense  objects such as core-collapse supernovae, compact binary mergers, and the early universe.  
 In such environments, neutrino-neutrino interactions induce non-linear effects in the flavor evolution, making the latter very challenging and counter-intuitive to grasp. A recent development in the field concerns the possible existence of fast flavor conversion stemming from the pairwise scattering of neutrinos.
 \item Existing analytical approaches employed to gauge the existence of instabilities in the flavor space  rely on the linear stability analysis. The numerical solution of the equations of motion is mainly challenged by the presence of characteristic frequencies differing  many orders of magnitude from each other.  
 \item  Fast pairwise conversion has received a lot of attention in the past few years because it may occur in the neutrino decoupling region possibly affecting the source physics, in addition to the observable neutrino signal. The growth of the flavor instability is mainly driven by the (anti)neutrino number density; hence,  the development of flavor conversions could occur on very small timescales. A peculiar aspect regarding fast pairwise conversion is that it does not  strongly depend on the neutrino energy distribution, like in the case of classical flavor conversion, but  on the angular distribution of the scattering neutrinos.
\item  Fast pairwise conversion is expected to develop when crossings occur in the effective electron lepton number angular distribution of (anti)neutrinos. Moreover, the non-linear regime of fast conversion  is strongly dependent on the neutrino vacuum frequency when the latter is not vanishingly small,  the strength of the electron neutrino lepton number crossings, and eventually collisions. 
\item  Favorable conditions for the occurrence of fast pairwise conversion exist  in core-collapse supernovae, with potential   implications on the explosion mechanism, as well as in compact binary mergers; however, preliminary work shows that the possible implications for the nucleosynthesis of the heavy elements in compact mergers (and  the kilonova lightcurve) appear to depend on the nature of the compact object at the center of the remnant disk, i.e.~whether the latter is a hypermassive neutron star or a black hole. The physics of fast pairwise conversion in the early universe has not been explored yet, despite the fact that local inhomogeneities  may lead to favorable conditions for the occurrence of pairwise conversion.
\end{enumerate}
\end{summary}

\begin{issues}[FUTURE ISSUES]
\begin{enumerate}
\item  {\it Progress on  conceptual understanding.} Recent work highlights intrinsic limitations of the predictive power of the linear stability analysis. A deeper understanding of fast pairwise conversion, jointly with  progress in numerical modeling, are necessary in order to finally grasp whether this phenomenon leads to large flavor mixing, as initially postulated. The validity of the mean field approximation should also be scrutinized in the light of our renewed understanding of neutrino self-interactions.
\item  {\it Phenomenology of flavor conversion when symmetry assumptions are relaxed.} In the context of slow neutrino conversion, it has been shown that the flavor phenomenology is strongly affected when symmetry assumptions are relaxed. Similar findings might hold for pairwise conversion and can further affect current conjectures.
\item  {\it Better understanding of flavor conversion physics.} It remains to be understood whether an interplay between fast and slow conversions or matter effects can occur and how this would affect the flavor outcome. The physics picture discussed here may also be drastically altered by non-standard physics.
\item {\it Feedback on  source physics.} If fast  conversion occurs in dense environments, leading to large flavor mixing, its coupling to the source physics may  be not negligible and it remains to  be explored.  
\end{enumerate}
\end{issues}

\section*{DISCLOSURE STATEMENT}
The authors are not aware of any affiliations, memberships, funding, or financial holdings that
might be perceived as affecting the objectivity of this review. 

\section*{ACKNOWLEDGMENTS}
We are grateful to Thomas Janka and Georg Raffelt for insightful discussions. This work was supported by the Villum Foundation (Project No.\ 13164),
the Carlsberg Foundation (Grant No.\ CF18-0183), the Danmarks Frie Forskningsfonds (Grant No.\ 8049-00038B), the Knud H{\o}jgaard Foundation, and the Deutsche Forschungsgemeinschaft through Grant No.\ SFB 1258 (Collaborative Research Center ``Neutrinos, Dark Matter, Messengers'').


\hoffset = -100pt
\oddsidemargin = 120pt
\textwidth = 400pt
\linewidth = \textwidth
\bibliographystyle{ar-style5.bst}
\bibliography{references}

\begin{thebibliography}{149}
\expandafter\ifx\csname natexlab\endcsname\relax\def\natexlab#1{#1}\fi

\bibitem{Vitagliano:2019yzm}
Vitagliano E, Tamborra I, Raffelt
  G\href{https://doi.org/10.1103/RevModPhys.92.045006}{.
\newblock \textit{Rev. Mod. Phys.} 92:45006} (2020)
  [\href{https://arxiv.org/abs/1910.11878}{arXiv:1910.11878}]

\bibitem{Duan:2010bg}
Duan H, Fuller GM, Qian
  YZ\href{https://doi.org/10.1146/annurev.nucl.012809.104524}{.
\newblock \textit{Ann. Rev. Nucl. Part. Sci.} 60:569} (2010)
  [\href{https://arxiv.org/abs/1001.2799}{arXiv:1001.2799}]

\bibitem{Mirizzi:2015eza}
Mirizzi A, et~al.\href{https://doi.org/10.1393/ncr/i2016-10120-8}{.
\newblock \textit{Riv. Nuovo Cim.} 39:1} (2016)
  [\href{https://arxiv.org/abs/1508.00785}{arXiv:1508.00785}]

\bibitem{Chakraborty:2016yeg}
Chakraborty S, Hansen R, Izaguirre I, Raffelt
  GG\href{https://doi.org/10.1016/j.nuclphysb.2016.02.012}{.
\newblock \textit{Nucl. Phys. B} 908:366} (2016)
  [\href{https://arxiv.org/abs/1602.02766}{arXiv:1602.02766}]

\bibitem{Mikheev:1986if}
Mikheev SP, Smirnov A{\relax Yu}.
\newblock \textit{Sov. Phys. JETP} 64:4 (1986)
  [\href{https://arxiv.org/abs/0706.0454}{arXiv:0706.0454}]

\bibitem{1985YaFiz..42.1441M}
{Mikheyev} SP, {Smirnov} A{\relax Yu}.
\newblock \textit{Yadernaya Fizika} 42:1441 (1985)

\bibitem{1978PhRvD..17.2369W}
{Wolfenstein} L\href{https://doi.org/10.1103/PhysRevD.17.2369}{.
\newblock \textit{Phys. Rev. D} 17:2369} (1978)

\bibitem{1987ApJ...322..795F}
{Fuller} GM, {Mayle} RW, {Wilson} JR, {Schramm}
  DN\href{https://doi.org/10.1086/165772}{.
\newblock \textit{Astrophys. J.} 322:795} (1987)

\bibitem{Notzold:1987ik}
N{\"o}tzold D, Raffelt GG\href{https://doi.org/10.1016/0550-3213(88)90113-7}{.
\newblock \textit{Nucl. Phys. B} 307:924} (1988)

\bibitem{Pantaleone:1992eq}
Pantaleone JT\href{https://doi.org/10.1016/0370-2693(92)91887-F}{.
\newblock \textit{Phys. Lett. B} 287:128} (1992)

\bibitem{Pantaleone:1992xh}
Pantaleone JT\href{https://doi.org/10.1103/PhysRevD.46.510}{.
\newblock \textit{Phys. Rev. D} 46:510} (1992)

\bibitem{Janka:2006fh}
Janka HT, et~al.\href{https://doi.org/10.1016/j.physrep.2007.02.002}{.
\newblock \textit{Phys. Rept.} 442:38} (2007)
  [\href{https://arxiv.org/abs/astro-ph/0612072}{arXiv:astro-ph/0612072}]

\bibitem{Janka:2017vcp}
Janka HT  .
\newblock {Neutrino-Driven Explosions, Handbook of Supernovae}.
\newblock  1095 (2017)

\bibitem{Burrows:2012ew}
Burrows A\href{https://doi.org/10.1103/RevModPhys.85.245}{.
\newblock \textit{Rev. Mod. Phys.} 85:245} (2013)
  [\href{https://arxiv.org/abs/1210.4921}{arXiv:1210.4921}]

\bibitem{Burrows:2020qrp}
Burrows A, Vartanyan D  (2020)
  [\href{https://arxiv.org/abs/2009.14157}{arXiv:2009.14157}]

\bibitem{Duan:2006an}
Duan H, Fuller GM, Carlson J, Qian
  YZ\href{https://doi.org/10.1103/PhysRevD.74.105014}{.
\newblock \textit{Phys. Rev. D} 74:105014} (2006)
  [\href{https://arxiv.org/abs/astro-ph/0606616}{arXiv:astro-ph/0606616}]

\bibitem{Fogli:2007bk}
Fogli G, Lisi E, Marrone A, Mirizzi
  A\href{https://doi.org/10.1088/1475-7516/2007/12/010}{.
\newblock \textit{JCAP} 12:010} (2007)
  [\href{https://arxiv.org/abs/0707.1998}{arXiv:0707.1998}]

\bibitem{Raffelt:2013rqa}
Raffelt GG, Sarikas S, de~Sousa~Seixas
  D\href{https://doi.org/10.1103/PhysRevLett.111.091101}{.
\newblock \textit{Phys. Rev. Lett.} 111:091101} (2013)
  [\href{https://arxiv.org/abs/1305.7140}{arXiv:1305.7140}] , [Erratum:
  Phys.Rev.Lett. 113, 239903 (2014)]

\bibitem{Duan:2014gfa}
Duan H, Shalgar S\href{https://doi.org/10.1016/j.physletb.2015.05.057}{.
\newblock \textit{Phys. Lett. B} 747:139} (2015)
  [\href{https://arxiv.org/abs/1412.7097}{arXiv:1412.7097}]

\bibitem{Abbar:2015mca}
Abbar S, Duan H, Shalgar S\href{https://doi.org/10.1103/PhysRevD.92.065019}{.
\newblock \textit{Phys. Rev. D} 92:065019} (2015)
  [\href{https://arxiv.org/abs/1507.08992}{arXiv:1507.08992}]

\bibitem{Mirizzi:2015fva}
Mirizzi A, Mangano G, Saviano
  N\href{https://doi.org/10.1103/PhysRevD.92.021702}{.
\newblock \textit{Phys. Rev. D} 92:021702} (2015)
  [\href{https://arxiv.org/abs/1503.03485}{arXiv:1503.03485}]

\bibitem{Cirigliano:2017hmk}
Cirigliano V, Paris MW, Shalgar
  S\href{https://doi.org/10.1016/j.physletb.2017.09.039}{.
\newblock \textit{Phys. Lett. B} 774:258} (2017)
  [\href{https://arxiv.org/abs/1706.07052}{arXiv:1706.07052}]

\bibitem{Sawyer:2005jk}
Sawyer RF\href{https://doi.org/10.1103/PhysRevD.72.045003}{.
\newblock \textit{Phys. Rev. D} 72:045003} (2005)
  [\href{https://arxiv.org/abs/hep-ph/0503013}{arXiv:hep-ph/0503013}]

\bibitem{Sawyer:2008zs}
Sawyer RF\href{https://doi.org/10.1103/PhysRevD.79.105003}{.
\newblock \textit{Phys. Rev.} D79:105003} (2009)
  [\href{https://arxiv.org/abs/0803.4319}{arXiv:0803.4319}]

\bibitem{Sawyer:2015dsa}
Sawyer RF\href{https://doi.org/10.1103/PhysRevLett.116.081101}{.
\newblock \textit{Phys. Rev. Lett.} 116:081101} (2016)
  [\href{https://arxiv.org/abs/1509.03323}{arXiv:1509.03323}]

\bibitem{Chakraborty:2016lct}
Chakraborty S, Hansen RS, Izaguirre I, Raffelt
  GG\href{https://doi.org/10.1088/1475-7516/2016/03/042}{.
\newblock \textit{JCAP} 03:042} (2016)
  [\href{https://arxiv.org/abs/1602.00698}{arXiv:1602.00698}]

\bibitem{Izaguirre:2016gsx}
Izaguirre I, Raffelt G, Tamborra
  I\href{https://doi.org/10.1103/PhysRevLett.118.021101}{.
\newblock \textit{Phys. Rev. Lett.} 118:021101} (2017)
  [\href{https://arxiv.org/abs/1610.01612}{arXiv:1610.01612}]

\bibitem{Friedland:2003eh}
Friedland A, Lunardini C\href{https://doi.org/10.1088/1126-6708/2003/10/043}{.
\newblock \textit{JHEP} 10:043} (2003)
  [\href{https://arxiv.org/abs/hep-ph/0307140}{arXiv:hep-ph/0307140}]

\bibitem{Friedland:2003dv}
Friedland A, Lunardini C\href{https://doi.org/10.1103/PhysRevD.68.013007}{.
\newblock \textit{Phys. Rev. D} 68:013007} (2003)
  [\href{https://arxiv.org/abs/hep-ph/0304055}{arXiv:hep-ph/0304055}]

\bibitem{Balantekin:2006tg}
Balantekin A, Pehlivan Y\href{https://doi.org/10.1088/0954-3899/34/1/004}{.
\newblock \textit{J. Phys. G} 34:47} (2007)
  [\href{https://arxiv.org/abs/astro-ph/0607527}{arXiv:astro-ph/0607527}]

\bibitem{Pehlivan:2010zz}
Pehlivan Y, et~al.\href{https://doi.org/10.1063/1.3485133}{.
\newblock \textit{AIP Conf. Proc.} 1269:189} (2010)

\bibitem{Pehlivan:2011hp}
Pehlivan Y, Balantekin A, Kajino T, Yoshida
  T\href{https://doi.org/10.1103/PhysRevD.84.065008}{.
\newblock \textit{Phys. Rev. D} 84:065008} (2011)
  [\href{https://arxiv.org/abs/1105.1182}{arXiv:1105.1182}]

\bibitem{Birol:2018qhx}
Birol S, Pehlivan Y, Balantekin A, Kajino
  T\href{https://doi.org/10.1103/PhysRevD.98.083002}{.
\newblock \textit{Phys. Rev. D} 98:083002} (2018)
  [\href{https://arxiv.org/abs/1805.11767}{arXiv:1805.11767}]

\bibitem{Patwardhan:2019zta}
Patwardhan AV, Cervia MJ, Baha~Balantekin
  A\href{https://doi.org/10.1103/PhysRevD.99.123013}{.
\newblock \textit{Phys. Rev. D} 99:123013} (2019)
  [\href{https://arxiv.org/abs/1905.04386}{arXiv:1905.04386}]

\bibitem{Cervia:2019res}
Cervia MJ, et~al.\href{https://doi.org/10.1103/PhysRevD.100.083001}{.
\newblock \textit{Phys. Rev. D} 100:083001} (2019)
  [\href{https://arxiv.org/abs/1908.03511}{arXiv:1908.03511}]

\bibitem{Rrapaj:2019pxz}
Rrapaj E\href{https://doi.org/10.1103/PhysRevC.101.065805}{.
\newblock \textit{Phys. Rev. C} 101:065805} (2020)
  [\href{https://arxiv.org/abs/1905.13335}{arXiv:1905.13335}]

\bibitem{Sigl:1992fn}
Sigl G, Raffelt GG\href{https://doi.org/10.1016/0550-3213(93)90175-O}{.
\newblock \textit{Nucl. Phys. B} 406:423} (1993)

\bibitem{Cardall:2007zw}
Cardall CY\href{https://doi.org/10.1103/PhysRevD.78.085017}{.
\newblock \textit{Phys. Rev. D} 78:085017} (2008)
  [\href{https://arxiv.org/abs/0712.1188}{arXiv:0712.1188}]

\bibitem{Stirner:2018ojk}
Stirner T, Sigl G, Raffelt
  GG\href{https://doi.org/10.1088/1475-7516/2018/05/016}{.
\newblock \textit{JCAP} 05:016} (2018)
  [\href{https://arxiv.org/abs/1803.04693}{arXiv:1803.04693}]

\bibitem{Nunokawa:1997ct}
Nunokawa H, Peltoniemi J, Rossi A, Valle
  J\href{https://doi.org/10.1103/PhysRevD.56.1704}{.
\newblock \textit{Phys. Rev. D} 56:1704} (1997)
  [\href{https://arxiv.org/abs/hep-ph/9702372}{arXiv:hep-ph/9702372}]

\bibitem{Stapleford:2016jgz}
Stapleford CJ, et~al.\href{https://doi.org/10.1103/PhysRevD.94.093007}{.
\newblock \textit{Phys. Rev. D} 94:093007} (2016)
  [\href{https://arxiv.org/abs/1605.04903}{arXiv:1605.04903}]

\bibitem{Wong:2002fa}
Wong YY\href{https://doi.org/10.1103/PhysRevD.66.025015}{.
\newblock \textit{Phys. Rev. D} 66:025015} (2002)
  [\href{https://arxiv.org/abs/hep-ph/0203180}{arXiv:hep-ph/0203180}]

\bibitem{Dolgov:2002ab}
Dolgov A, et~al.\href{https://doi.org/10.1016/S0550-3213(02)00274-2}{.
\newblock \textit{Nucl. Phys. B} 632:363} (2002)
  [\href{https://arxiv.org/abs/hep-ph/0201287}{arXiv:hep-ph/0201287}]

\bibitem{Johns:2016enc}
Johns L, et~al.\href{https://doi.org/10.1103/PhysRevD.94.083505}{.
\newblock \textit{Phys. Rev. D} 94:083505} (2016)
  [\href{https://arxiv.org/abs/1608.01336}{arXiv:1608.01336}]

\bibitem{Banerjee:2011fj}
Banerjee A, Dighe A, Raffelt
  GG\href{https://doi.org/10.1103/PhysRevD.84.053013}{.
\newblock \textit{Phys. Rev. D} 84:053013} (2011)
  [\href{https://arxiv.org/abs/1107.2308}{arXiv:1107.2308}]

\bibitem{Sarikas:2011am}
Sarikas S, Raffelt GG, H{\"u}depohl L, Janka
  HT\href{https://doi.org/10.1103/PhysRevLett.108.061101}{.
\newblock \textit{Phys. Rev. Lett.} 108:061101} (2012)
  [\href{https://arxiv.org/abs/1109.3601}{arXiv:1109.3601}]

\bibitem{Abbar:2017pkh}
Abbar S, Duan H\href{https://doi.org/10.1103/PhysRevD.98.043014}{.
\newblock \textit{Phys. Rev. D} 98:043014} (2018)
  [\href{https://arxiv.org/abs/1712.07013}{arXiv:1712.07013}]

\bibitem{Shalgar:2019qwg}
Shalgar S, Padilla-Gay I, Tamborra
  I\href{https://doi.org/10.1088/1475-7516/2020/06/048}{.
\newblock \textit{JCAP} 06:048} (2020)
  [\href{https://arxiv.org/abs/1911.09110}{arXiv:1911.09110}]

\bibitem{Capozzi:2019lso}
Capozzi F, Raffelt GG, Stirner
  T\href{https://doi.org/10.1088/1475-7516/2019/09/002}{.
\newblock \textit{JCAP} 09:002} (2019)
  [\href{https://arxiv.org/abs/1906.08794}{arXiv:1906.08794}]

\bibitem{Stirner:2020}
Stirner T. 2020.
\newblock Fast neutrino flavour conversions.
\newblock {PhD} thesis, Ludwig-Maximilians-Universit{\"a}t M{\"u}nchen

\bibitem{Capozzi:2017gqd}
Capozzi F, et~al.\href{https://doi.org/10.1103/PhysRevD.96.043016}{.
\newblock \textit{Phys. Rev. D} 96:043016} (2017)
  [\href{https://arxiv.org/abs/1706.03360}{arXiv:1706.03360}]

\bibitem{Sturrock:1958zz}
Sturrock P\href{https://doi.org/10.1103/PhysRev.112.1488}{.
\newblock \textit{Phys. Rev.} 112:1488} (1958)

\bibitem{Yi:2019hrp}
Yi C, Ma L, Martin JD, Duan
  H\href{https://doi.org/10.1103/PhysRevD.99.063005}{.
\newblock \textit{Phys. Rev. D} 99:063005} (2019)
  [\href{https://arxiv.org/abs/1901.01546}{arXiv:1901.01546}]

\bibitem{Shalgar:2019rqe}
Shalgar S, Tamborra I, Bustamante M  (2019)
  [\href{https://arxiv.org/abs/1912.09115}{arXiv:1912.09115}]

\bibitem{Chakraborty:2019wxe}
Chakraborty M, Chakraborty
  S\href{https://doi.org/10.1088/1475-7516/2020/01/005}{.
\newblock \textit{JCAP} 01:005} (2020)
  [\href{https://arxiv.org/abs/1909.10420}{arXiv:1909.10420}]

\bibitem{Capozzi:2020kge}
Capozzi F, Chakraborty M, Chakraborty S, Sen
  M\href{https://doi.org/10.1103/PhysRevLett.125.251801}{.
\newblock \textit{Phys. Rev. Lett.} 125:251801} (2020)
  [\href{https://arxiv.org/abs/2005.14204}{arXiv:2005.14204}]

\bibitem{Dighe:2017sur}
Dighe A, Sen M\href{https://doi.org/10.1103/PhysRevD.97.043011}{.
\newblock \textit{Phys. Rev. D} 97:043011} (2018)
  [\href{https://arxiv.org/abs/1709.06858}{arXiv:1709.06858}]

\bibitem{EstebanPretel:2008ni}
Esteban-Pretel A, et~al.\href{https://doi.org/10.1103/PhysRevD.78.085012}{.
\newblock \textit{Phys. Rev. D} 78:085012} (2008)
  [\href{https://arxiv.org/abs/0807.0659}{arXiv:0807.0659}]

\bibitem{Airen:2018nvp}
Airen S, et~al.\href{https://doi.org/10.1088/1475-7516/2018/12/019}{.
\newblock \textit{JCAP} 12:019} (2018)
  [\href{https://arxiv.org/abs/1809.09137}{arXiv:1809.09137}]

\bibitem{Dasgupta:2017oko}
Dasgupta B, Sen M\href{https://doi.org/10.1103/PhysRevD.97.023017}{.
\newblock \textit{Phys. Rev. D} 97:023017} (2018)
  [\href{https://arxiv.org/abs/1709.08671}{arXiv:1709.08671}]

\bibitem{Johns:2019izj}
Johns L, Nagakura H, Fuller GM, Burrows
  A\href{https://doi.org/10.1103/PhysRevD.101.043009}{.
\newblock \textit{Phys. Rev. D} 101:043009} (2020)
  [\href{https://arxiv.org/abs/1910.05682}{arXiv:1910.05682}]

\bibitem{Shalgar:2020xns}
Shalgar S, Tamborra I\href{https://doi.org/10.1088/1475-7516/2021/01/014}{.
\newblock \textit{JCAP} 01:014} (2021)
  [\href{https://arxiv.org/abs/2007.07926}{arXiv:2007.07926}]

\bibitem{Hannestad:2006nj}
Hannestad S, Raffelt GG, Sigl G, Wong
  YY\href{https://doi.org/10.1103/PhysRevD.74.105010}{.
\newblock \textit{Phys. Rev. D} 74:105010} (2006)
  [\href{https://arxiv.org/abs/astro-ph/0608695}{arXiv:astro-ph/0608695}] ,
  [Erratum: Phys.Rev.D 76, 029901 (2007)]

\bibitem{Martin:2019gxb}
Martin JD, Yi C, Duan H\href{https://doi.org/10.1016/j.physletb.2019.135088}{.
\newblock \textit{Phys. Lett. B} 800:135088} (2020)
  [\href{https://arxiv.org/abs/1909.05225}{arXiv:1909.05225}]

\bibitem{Capozzi:2018clo}
Capozzi F, et~al.\href{https://doi.org/10.1103/PhysRevLett.122.091101}{.
\newblock \textit{Phys. Rev. Lett.} 122:091101} (2019)
  [\href{https://arxiv.org/abs/1808.06618}{arXiv:1808.06618}]

\bibitem{Shalgar:2019kzy}
Shalgar S, Tamborra I\href{https://doi.org/10.3847/1538-4357/ab38ba}{.
\newblock \textit{Astrophys. J.} 883:80} (2019)
  [\href{https://arxiv.org/abs/1904.07236}{arXiv:1904.07236}]

\bibitem{Shalgar:2020wcx}
Shalgar S, Tamborra I  (2020)
  [\href{https://arxiv.org/abs/2011.00004}{arXiv:2011.00004}]

\bibitem{Brandt:2010xa}
Brandt TD, Burrows A, Ott CD, Livne
  E\href{https://doi.org/10.1088/0004-637X/728/1/8}{.
\newblock \textit{Astrophys. J.} 728:8} (2011)
  [\href{https://arxiv.org/abs/1009.4654}{arXiv:1009.4654}]

\bibitem{Tamborra:2017ubu}
Tamborra I, H{\"u}edepohl L, Raffelt GG, Janka
  HT\href{https://doi.org/10.3847/1538-4357/aa6a18}{.
\newblock \textit{Astrophys. J.} 839:132} (2017)
  [\href{https://arxiv.org/abs/1702.00060}{arXiv:1702.00060}]

\bibitem{Tamborra:2014aua}
Tamborra I, et~al.\href{https://doi.org/10.1088/0004-637X/792/2/96}{.
\newblock \textit{Astrophys. J.} 792:96} (2014)
  [\href{https://arxiv.org/abs/1402.5418}{arXiv:1402.5418}]

\bibitem{Abbar:2020ror}
Abbar S  (2020) [\href{https://arxiv.org/abs/2007.13655}{arXiv:2007.13655}]

\bibitem{Sarikas:2012ad}
Sarikas S, de~Sousa~Seixas D, Raffelt
  GG\href{https://doi.org/10.1103/PhysRevD.86.125020}{.
\newblock \textit{Phys. Rev. D} 86:125020} (2012)
  [\href{https://arxiv.org/abs/1210.4557}{arXiv:1210.4557}]

\bibitem{Morinaga:2018aug}
Morinaga T, Yamada S\href{https://doi.org/10.1103/PhysRevD.97.023024}{.
\newblock \textit{Phys. Rev. D} 97:023024} (2018)
  [\href{https://arxiv.org/abs/1803.05913}{arXiv:1803.05913}]

\bibitem{Johns:2020qsk}
Johns L, Nagakura H, Fuller GM, Burrows
  A\href{https://doi.org/10.1103/PhysRevD.102.103017}{.
\newblock \textit{Phys. Rev. D} 102:103017} (2020)
  [\href{https://arxiv.org/abs/2009.09024}{arXiv:2009.09024}]

\bibitem{Bhattacharyya:2020jpj}
Bhattacharyya S, Dasgupta B  (2020)
  [\href{https://arxiv.org/abs/2009.03337}{arXiv:2009.03337}]

\bibitem{Bethe:1984ux}
Bethe HA, Wilson JR\href{https://doi.org/10.1086/163343}{.
\newblock \textit{Astrophys. J.} 295:14} (1985)

\bibitem{shapiro2008black}
Shapiro SL, Teukolsky SA.
\newblock John Wiley \& Sons (2008)

\bibitem{Bollig:2017lki}
Bollig R, et~al.\href{https://doi.org/10.1103/PhysRevLett.119.242702}{.
\newblock \textit{Phys. Rev. Lett.} 119:242702} (2017)
  [\href{https://arxiv.org/abs/1706.04630}{arXiv:1706.04630}]

\bibitem{Janka:2016fox}
Janka HT, Melson T, Summa
  A\href{https://doi.org/10.1146/annurev-nucl-102115-044747}{.
\newblock \textit{Ann. Rev. Nucl. Part. Sci.} 66:341} (2016)
  [\href{https://arxiv.org/abs/1602.05576}{arXiv:1602.05576}]

\bibitem{Burrows:2019xff}
Burrows A. 2019.
\newblock In \textit{{International Conference on History of the Neutrino}:
  {1930-2018}}

\bibitem{Mezzacappa:2020oyq}
Mezzacappa A, Endeve E, Messer OB, Bruenn SW  (2020)
  [\href{https://arxiv.org/abs/2010.09013}{arXiv:2010.09013}]

\bibitem{Vartanyan:2019ssu}
Vartanyan D, Burrows A, Radice D\href{https://doi.org/10.1093/mnras/stz2307}{.
\newblock \textit{Mon. Not. Roy. Astron. Soc.} 489:2227} (2019)
  [\href{https://arxiv.org/abs/1906.08787}{arXiv:1906.08787}]

\bibitem{Nagakura:2017mnp}
Nagakura H, et~al.\href{https://doi.org/10.3847/1538-4357/aaac29}{.
\newblock \textit{Astrophys. J.} 854:136} (2018)
  [\href{https://arxiv.org/abs/1702.01752}{arXiv:1702.01752}]

\bibitem{Bruenn:2014qea}
Bruenn SW, et~al.\href{https://doi.org/10.3847/0004-637X/818/2/123}{.
\newblock \textit{Astrophys. J.} 818:123} (2016)
  [\href{https://arxiv.org/abs/1409.5779}{arXiv:1409.5779}]

\bibitem{OConnor:2015rwy}
O'Connor EP, Couch SM\href{https://doi.org/10.3847/1538-4357/aaa893}{.
\newblock \textit{Astrophys. J.} 854:63} (2018)
  [\href{https://arxiv.org/abs/1511.07443}{arXiv:1511.07443}]

\bibitem{Richers:2017awc}
Richers S, et~al.\href{https://doi.org/10.3847/1538-4357/aa8bb2}{.
\newblock \textit{Astrophys. J.} 847:133} (2017)
  [\href{https://arxiv.org/abs/1706.06187}{arXiv:1706.06187}]

\bibitem{Vartanyan:2018xcd}
Vartanyan D, et~al.\href{https://doi.org/10.1093/mnras/sty809}{.
\newblock \textit{Mon. Not. Roy. Astron. Soc.} 477:3091} (2018)
  [\href{https://arxiv.org/abs/1801.08148}{arXiv:1801.08148}]

\bibitem{Just:2018djz}
Just O, et~al.\href{https://doi.org/10.1093/mnras/sty2578}{.
\newblock \textit{Mon. Not. Roy. Astron. Soc.} 481:4786} (2018)
  [\href{https://arxiv.org/abs/1805.03953}{arXiv:1805.03953}]

\bibitem{Cabezon:2018lpr}
Cabez{\'o}n RM, et~al.\href{https://doi.org/10.1051/0004-6361/201833705}{.
\newblock \textit{Astron. Astrophys.} 619:A118} (2018)
  [\href{https://arxiv.org/abs/1806.09184}{arXiv:1806.09184}]

\bibitem{Pan:2018vkx}
Pan KC, et~al.\href{https://doi.org/10.1088/1361-6471/aaed51}{.
\newblock \textit{J. Phys. G} 46:014001} (2019)
  [\href{https://arxiv.org/abs/1806.10030}{arXiv:1806.10030}]

\bibitem{Glas:2018vcs}
{Glas} R, et~al.\href{https://doi.org/10.3847/1538-4357/ab275c}{.
\newblock \textit{Astrophys. J.} 881:36} (2019)
  [\href{https://arxiv.org/abs/1809.10150}{arXiv:1809.10150}]

\bibitem{Glas:2018oyz}
Glas R, Just O, Janka HT, Obergaulinger
  M\href{https://doi.org/10.3847/1538-4357/ab0423}{.
\newblock \textit{Astrophys. J.} 873:45} (2019)
  [\href{https://arxiv.org/abs/1809.10146}{arXiv:1809.10146}]

\bibitem{Dasgupta:2018ulw}
Dasgupta B, Mirizzi A, Sen M\href{https://doi.org/10.1103/PhysRevD.98.103001}{.
\newblock \textit{Phys. Rev. D} 98:103001} (2018)
  [\href{https://arxiv.org/abs/1807.03322}{arXiv:1807.03322}]

\bibitem{Abbar:2020fcl}
Abbar S\href{https://doi.org/10.1088/1475-7516/2020/05/027}{.
\newblock \textit{JCAP} 05:027} (2020)
  [\href{https://arxiv.org/abs/2003.00969}{arXiv:2003.00969}]

\bibitem{Abbar:2018shq}
Abbar S, et~al.\href{https://doi.org/10.1103/PhysRevD.100.043004}{.
\newblock \textit{Phys. Rev. D} 100:043004} (2019)
  [\href{https://arxiv.org/abs/1812.06883}{arXiv:1812.06883}]

\bibitem{Azari:2019jvr}
Delfan~Azari M, et~al.\href{https://doi.org/10.1103/PhysRevD.99.103011}{.
\newblock \textit{Phys. Rev. D} 99:103011} (2019)
  [\href{https://arxiv.org/abs/1902.07467}{arXiv:1902.07467}]

\bibitem{DelfanAzari:2019tez}
Delfan~Azari M, et~al.\href{https://doi.org/10.1103/PhysRevD.101.023018}{.
\newblock \textit{Phys. Rev. D} 101:023018} (2020)
  [\href{https://arxiv.org/abs/1910.06176}{arXiv:1910.06176}]

\bibitem{Glas:2019ijo}
Glas R, et~al.\href{https://doi.org/10.1103/PhysRevD.101.063001}{.
\newblock \textit{Phys. Rev. D} 101:063001} (2020)
  [\href{https://arxiv.org/abs/1912.00274}{arXiv:1912.00274}]

\bibitem{Abbar:2019zoq}
Abbar S, et~al.\href{https://doi.org/10.1103/PhysRevD.101.043016}{.
\newblock \textit{Phys. Rev. D} 101:043016} (2020)
  [\href{https://arxiv.org/abs/1911.01983}{arXiv:1911.01983}]

\bibitem{Nagakura:2019sig}
{Nagakura} H, {Morinaga} T, {Kato} C, {Yamada}
  S\href{https://doi.org/10.3847/1538-4357/ab4cf2}{.
\newblock \textit{Astrophys. J.} 886:139} (2019)
  [\href{https://arxiv.org/abs/1910.04288}{arXiv:1910.04288}]

\bibitem{Morinaga:2019wsv}
Morinaga T, Nagakura H, Kato C, Yamada
  S\href{https://doi.org/10.1103/PhysRevResearch.2.012046}{.
\newblock \textit{Phys. Rev. Res.} 2:012046} (2020)
  [\href{https://arxiv.org/abs/1909.13131}{arXiv:1909.13131}]

\bibitem{Xiong:2020ntn}
Xiong Z, Sieverding A, Sen M, Qian YZ  (2020)
  [\href{https://arxiv.org/abs/2006.11414}{arXiv:2006.11414}]

\bibitem{Ruffert:1996by}
Ruffert M, Janka HT, Takahashi K, Sch{\"a}fer G.
\newblock \textit{Astron. Astrophys.} 319:122 (1997)
  [\href{https://arxiv.org/abs/astro-ph/9606181}{arXiv:astro-ph/9606181}]

\bibitem{Foucart:2015vpa}
Foucart F, et~al.\href{https://doi.org/10.1103/PhysRevD.91.124021}{.
\newblock \textit{Phys. Rev. D} 91:124021} (2015)
  [\href{https://arxiv.org/abs/1502.04146}{arXiv:1502.04146}]

\bibitem{Abe:2018mic}
Hayato Y, et~al.\href{https://doi.org/10.3847/2041-8213/aabaca}{.
\newblock \textit{Astrophys. J. Lett.} 857:L4} (2018)
  [\href{https://arxiv.org/abs/1802.04379}{arXiv:1802.04379}]

\bibitem{1974ApJ192L.145L}
{Lattimer} JM, {Schramm} DN\href{https://doi.org/10.1086/181612}{.
\newblock \textit{Astropys. J. Lett.} 192:L145} (1974)

\bibitem{Eichler:1989ve}
Eichler D, Livio M, Piran T, Schramm
  DN\href{https://doi.org/10.1038/340126a0}{.
\newblock \textit{Nature} 340:126} (1989)

\bibitem{Li:1998bw}
Li LX, Paczynski B\href{https://doi.org/10.1086/311680}{.
\newblock \textit{Astrophys. J. Lett.} 507:L59} (1998)
  [\href{https://arxiv.org/abs/astro-ph/9807272}{arXiv:astro-ph/9807272}]

\bibitem{Kulkarni:2005jw}
Kulkarni SR  (2005)
  [\href{https://arxiv.org/abs/astro-ph/0510256}{arXiv:astro-ph/0510256}]

\bibitem{Metzger:2010sy}
Metzger B, et~al.\href{https://doi.org/10.1111/j.1365-2966.2010.16864.x}{.
\newblock \textit{Mon. Not. Roy. Astron. Soc.} 406:2650} (2010)
  [\href{https://arxiv.org/abs/1001.5029}{arXiv:1001.5029}]

\bibitem{Wanajo:2014wha}
Wanajo S, et~al.\href{https://doi.org/10.1088/2041-8205/789/2/L39}{.
\newblock \textit{Astrophys. J. Lett.} 789:L39} (2014)
  [\href{https://arxiv.org/abs/1402.7317}{arXiv:1402.7317}]

\bibitem{Perego:2014fma}
Perego A, et~al.\href{https://doi.org/10.1093/mnras/stu1352}{.
\newblock \textit{Mon. Not. Roy. Astron. Soc.} 443:3134} (2014)
  [\href{https://arxiv.org/abs/1405.6730}{arXiv:1405.6730}]

\bibitem{Fernandez:2015use}
Fern{\'a}ndez R, Metzger
  BD\href{https://doi.org/10.1146/annurev-nucl-102115-044819}{.
\newblock \textit{Ann. Rev. Nucl. Part. Sci.} 66:23} (2016)
  [\href{https://arxiv.org/abs/1512.05435}{arXiv:1512.05435}]

\bibitem{Sekiguchi:2015dma}
Sekiguchi Y, Kiuchi K, Kyutoku K, Shibata
  M\href{https://doi.org/10.1103/PhysRevD.91.064059}{.
\newblock \textit{Phys. Rev. D} 91:064059} (2015)
  [\href{https://arxiv.org/abs/1502.06660}{arXiv:1502.06660}]

\bibitem{Radice:2016dwd}
Radice D, et~al.\href{https://doi.org/10.1093/mnras/stw1227}{.
\newblock \textit{Mon. Not. Roy. Astron. Soc.} 460:3255} (2016)
  [\href{https://arxiv.org/abs/1601.02426}{arXiv:1601.02426}]

\bibitem{Miller:2019dpt}
Miller JM, et~al.\href{https://doi.org/10.1103/PhysRevD.100.023008}{.
\newblock \textit{Phys. Rev. D} 100:023008} (2019)
  [\href{https://arxiv.org/abs/1905.07477}{arXiv:1905.07477}]

\bibitem{Woosley:1993wj}
Woosley SE\href{https://doi.org/10.1086/172359}{.
\newblock \textit{Astrophys. J.} 405:273} (1993)

\bibitem{Ruffert:1998qg}
Ruffert M, Janka HT.
\newblock \textit{Astron. Astrophys.} 344:573 (1999)
  [\href{https://arxiv.org/abs/astro-ph/9809280}{arXiv:astro-ph/9809280}]

\bibitem{2011MNRAS.410.2302Z}
{Zalamea} I, {Beloborodov}
  AM\href{https://doi.org/10.1111/j.1365-2966.2010.17600.x}{.
\newblock \textit{Mon. Not. Roy. Astron. Soc.} 410:2302} (2011)
  [\href{https://arxiv.org/abs/1003.0710}{arXiv:1003.0710}]

\bibitem{Just:2015dba}
Just O, et~al.\href{https://doi.org/10.3847/2041-8205/816/2/L30}{.
\newblock \textit{Astrophys. J. Lett.} 816:L30} (2016)
  [\href{https://arxiv.org/abs/1510.04288}{arXiv:1510.04288}]

\bibitem{Foucart:2020qjb}
Foucart F, et~al.\href{https://doi.org/10.3847/2041-8213/abbb87}{.
\newblock \textit{Astrophys. J. Lett.} 902:L27} (2020)
  [\href{https://arxiv.org/abs/2008.08089}{arXiv:2008.08089}]

\bibitem{TheLIGOScientific:2017qsa}
Abbott B, et~al.\href{https://doi.org/10.1103/PhysRevLett.119.161101}{.
\newblock \textit{Phys. Rev. Lett.} 119:161101} (2017)
  [\href{https://arxiv.org/abs/1710.05832}{arXiv:1710.05832}]

\bibitem{Monitor:2017mdv}
Abbott B, et~al.\href{https://doi.org/10.3847/2041-8213/aa920c}{.
\newblock \textit{Astrophys. J. Lett.} 848:L13} (2017)
  [\href{https://arxiv.org/abs/1710.05834}{arXiv:1710.05834}]

\bibitem{GBM:2017lvd}
Abbott B, et~al.\href{https://doi.org/10.3847/2041-8213/aa91c9}{.
\newblock \textit{Astrophys. J. Lett.} 848:L12} (2017)
  [\href{https://arxiv.org/abs/1710.05833}{arXiv:1710.05833}]

\bibitem{Foucart:2014nda}
Foucart F, et~al.\href{https://doi.org/10.1103/PhysRevD.90.024026}{.
\newblock \textit{Phys. Rev. D} 90:024026} (2014)
  [\href{https://arxiv.org/abs/1405.1121}{arXiv:1405.1121}]

\bibitem{Just:2014fka}
Just O, et~al.\href{https://doi.org/10.1093/mnras/stv009}{.
\newblock \textit{Mon. Not. Roy. Astron. Soc.} 448:541} (2015)
  [\href{https://arxiv.org/abs/1406.2687}{arXiv:1406.2687}]

\bibitem{Fernandez:2013tya}
Fern{\'a}ndez R, Metzger BD\href{https://doi.org/10.1093/mnras/stt1312}{.
\newblock \textit{Mon. Not. Roy. Astron. Soc.} 435:502} (2013)
  [\href{https://arxiv.org/abs/1304.6720}{arXiv:1304.6720}]

\bibitem{Lippuner:2017bfm}
Lippuner J, et~al.\href{https://doi.org/10.1093/mnras/stx1987}{.
\newblock \textit{Mon. Not. Roy. Astron. Soc.} 472:904} (2017)
  [\href{https://arxiv.org/abs/1703.06216}{arXiv:1703.06216}]

\bibitem{Fujibayashi:2017puw}
Fujibayashi S, et~al.\href{https://doi.org/10.3847/1538-4357/aabafd}{.
\newblock \textit{Astrophys. J.} 860:64} (2018)
  [\href{https://arxiv.org/abs/1711.02093}{arXiv:1711.02093}]

\bibitem{Wu:2017drk}
Wu MR, Tamborra I, Just O, Janka
  HT\href{https://doi.org/10.1103/PhysRevD.96.123015}{.
\newblock \textit{Phys. Rev.} D96:123015} (2017)
  [\href{https://arxiv.org/abs/1711.00477}{arXiv:1711.00477}]

\bibitem{Ardevol-Pulpillo:2018btx}
Ardevol-Pulpillo R, Janka HT, Just O, Bauswein
  A\href{https://doi.org/10.1093/mnras/stz613}{.
\newblock \textit{Mon. Not. Roy. Astron. Soc.} 485:4754} (2019)
  [\href{https://arxiv.org/abs/1808.00006}{arXiv:1808.00006}]

\bibitem{George:2020veu}
George M, et~al.\href{https://doi.org/10.1103/PhysRevD.102.103015}{.
\newblock \textit{Phys. Rev. D} 102:103015} (2020)
  [\href{https://arxiv.org/abs/2009.04046}{arXiv:2009.04046}]

\bibitem{Malkus:2014iqa}
Malkus A, Friedland A, McLaughlin G  (2014)
  [\href{https://arxiv.org/abs/1403.5797}{arXiv:1403.5797}]

\bibitem{Malkus:2012ts}
Malkus A, Kneller J, McLaughlin G, Surman
  R\href{https://doi.org/10.1103/PhysRevD.86.085015}{.
\newblock \textit{Phys. Rev. D} 86:085015} (2012)
  [\href{https://arxiv.org/abs/1207.6648}{arXiv:1207.6648}]

\bibitem{Wu:2015fga}
Wu MR, Duan H, Qian YZ\href{https://doi.org/10.1016/j.physletb.2015.11.027}{.
\newblock \textit{Phys. Lett. B} 752:89} (2016)
  [\href{https://arxiv.org/abs/1509.08975}{arXiv:1509.08975}]

\bibitem{Tian:2017xbr}
Tian JY, Patwardhan AV, Fuller
  GM\href{https://doi.org/10.1103/PhysRevD.96.043001}{.
\newblock \textit{Phys. Rev. D} 96:043001} (2017)
  [\href{https://arxiv.org/abs/1703.03039}{arXiv:1703.03039}]

\bibitem{Vlasenko:2018irq}
Vlasenko A, McLaughlin G\href{https://doi.org/10.1103/PhysRevD.97.083011}{.
\newblock \textit{Phys. Rev. D} 97:083011} (2018)
  [\href{https://arxiv.org/abs/1801.07813}{arXiv:1801.07813}]

\bibitem{Shalgar:2017pzd}
Shalgar S\href{https://doi.org/10.1088/1475-7516/2018/02/010}{.
\newblock \textit{JCAP} 02:010} (2018)
  [\href{https://arxiv.org/abs/1707.07692}{arXiv:1707.07692}]

\bibitem{Wu:2017qpc}
Wu MR, Tamborra I\href{https://doi.org/10.1103/PhysRevD.95.103007}{.
\newblock \textit{Phys. Rev.} D95:103007} (2017)
  [\href{https://arxiv.org/abs/1701.06580}{arXiv:1701.06580}]

\bibitem{Padilla-Gay:2020uxa}
Padilla-Gay I, Shalgar S, Tamborra
  I\href{https://doi.org/10.1088/1475-7516/2021/01/017}{.
\newblock \textit{JCAP} 01:017} (2021)
  [\href{https://arxiv.org/abs/2009.01843}{arXiv:2009.01843}]

\bibitem{lesgourgues2013neutrino}
Lesgourgues J, Mangano G, Miele G, Pastor S.
\newblock Cambridge University Press (2013)

\bibitem{Zyla:2020zbs}
Zyla P, et~al.\href{https://doi.org/10.1093/ptep/ptaa104}{.
\newblock \textit{PTEP} 2020:083C01} (2020)

\bibitem{Akita:2020szl}
Akita K, Yamaguchi M\href{https://doi.org/10.1088/1475-7516/2020/08/012}{.
\newblock \textit{JCAP} 08:012} (2020)
  [\href{https://arxiv.org/abs/2005.07047}{arXiv:2005.07047}]

\bibitem{Lunardini:2001pb}
Lunardini C, Smirnov A\href{https://doi.org/10.1016/S0550-3213(01)00468-0}{.
\newblock \textit{Nucl. Phys. B} 616:307} (2001)
  [\href{https://arxiv.org/abs/hep-ph/0106149}{arXiv:hep-ph/0106149}]

\bibitem{Abazajian:2002qx}
Abazajian KN, Beacom JF, Bell
  NF\href{https://doi.org/10.1103/PhysRevD.66.013008}{.
\newblock \textit{Phys. Rev. D} 66:013008} (2002)
  [\href{https://arxiv.org/abs/astro-ph/0203442}{arXiv:astro-ph/0203442}]

\bibitem{Pastor:2001iu}
Pastor S, Raffelt GG, Semikoz
  DV\href{https://doi.org/10.1103/PhysRevD.65.053011}{.
\newblock \textit{Phys. Rev. D} 65:053011} (2002)
  [\href{https://arxiv.org/abs/hep-ph/0109035}{arXiv:hep-ph/0109035}]

\bibitem{Froustey:2020mcq}
Froustey J, Pitrou C, Volpe
  MC\href{https://doi.org/10.1088/1475-7516/2020/12/015}{.
\newblock \textit{JCAP} 12:015} (2020)
  [\href{https://arxiv.org/abs/2008.01074}{arXiv:2008.01074}]

\bibitem{Hansen:2020vgm}
Hansen RS, Shalgar S, Tamborra I  (2020)
  [\href{https://arxiv.org/abs/2012.03948}{arXiv:2012.03948}]

\bibitem{Bell:1998ds}
Bell NF, Volkas RR, Wong YY\href{https://doi.org/10.1103/PhysRevD.59.113001}{.
\newblock \textit{Phys. Rev. D} 59:113001} (1999)
  [\href{https://arxiv.org/abs/hep-ph/9809363}{arXiv:hep-ph/9809363}]

\end{thebibliography}


\end{document}